\documentstyle[12pt,fleqn]{article}
\catcode`\@=11
\@addtoreset{equation}{section}
\def\ksection{\arabic{section}}
\def\theequation{\ksection.\arabic{equation}}

\def\r{{r}}
\def\p{{\xi}}
\def\s{{s}}
\def\t{{t}}
\def\ti{{\t_{\rm i}}}
\def\DR{{\Delta R(\ti)}}

\def\rmx{{R_B}}
\def\Rmx{{R_B}}

\def\RW{{R_{\rm wall}(\ti)}}
\def\rw{{\Delta R_{\rm wall}(\ti)}}
\def\rwt{{\Delta R_{\rm wall}}}
\def\Hm{{H^{-1}_{\rm out}(\ti)}}
\def\x{{(R(\ti)-\RW)/\rw}}
\def\rtw{{\widetilde{r}}}

\def\ttw{{\widetilde{t}}}
\def\ep{{\epsilon}}

\def\gam{{\gamma}}

\def\alp{{\alpha}}
\def\bet{{\beta}}

\def\albe{{\alp\bet}}

\def\Lam{{\Lambda}}
\def\f{{\overline{f}}}

\def\rs{{\dot{r}_s}}

\def\dr{{\Delta R(\ti)}}

\def\dLam{{\dot{\Lam}}}
\def\dG{{\dot{\Gamma}}}
\def\dl{{\dot{l}}}
\def\dn{{\dot{n}}}
\def\dR{{\dot{R}}}
\def\dM{{\dot{M}}}
\def\dU{{\dot{U}}}
\def\drho{{\dot{\rho}}}
\def\dep{{\dot{\ep}}}
\def\der{{\nabla}}

\def\be{\begin{equation}}
\def\ee{\end{equation}}
\def\ba{\begin{eqnarray}}
\def\ea{\end{eqnarray}}
\def\la{\mathrel{\mathpalette\fun <}}
\def\ga{\mathrel{\mathpalette\fun >}}
\def\fun#1#2{\lower3.6pt\vbox{\baselineskip0pt\lineskip.9pt
        \ialign{$\mathsurround=0pt#1\hfill##\hfil$\crcr#2\crcr\sim\crcr}}}

\begin{document}
\begin{titlepage}
\null\vspace{-62pt}
\renewcommand{\thefootnote}{\alph{footnote}}
\begin{flushright}FNAL--PUB--93/005-A\\
March, 1993
\end{flushright}
\vspace{0.5in}
\centerline{{\bf{NUMERICAL EVOLUTION OF GENERAL RELATIVISTIC
VOIDS}}\footnote{Presented
as a thesis to the Department of Physics, The University of Chicago, in
partial fulfillment of the requirements for the Ph.D. degree}}
\vspace{.35in}
\centerline{Sharon L.~Vadas}
\vspace{.35in}
\centerline{{\it Physics Dept.~and Enrico Fermi Institute, The University of
Chicago, Chicago,
IL  60637}}
\centerline{{ \it and NASA/Fermilab Astrophysics Center}}
\centerline{{\it Fermi National Accelerator Laboratory, Batavia, IL  60510}}
\vspace{.4in}
\centerline{\bf Abstract}
\baselineskip=12pt
\begin{quotation}

In this paper, we study the evolution of a relativistic,
superhorizon-sized void
embedded in a Friedmann-Robertson-Walker universe.
We numerically solve the spherically symmetric general relativistic
equations in comoving, synchronous coordinates.
Initially, the fluid inside the void is taken to be homogeneous
and nonexpanding.
In a radiation-dominated universe, we find that radiation diffuses into the
void at approximately the speed of light
as a strong shock---the void collapses.
We also find the surprising result that the cosmic collapse time
(the $1^{\rm st}$-crossing time)
is much smaller than previously thought, because it
depends not only on the radius of the void, but also on the
ratio of the temperature inside the void to that outside.
If the ratio of the initial void radius to the outside Hubble radius
is less than the ratio of the outside temperature to that inside,
then the collapse occurs in
less than the outside Hubble time.
Thus, superhorizon-sized relativistic voids may
thermalize and homogenize relatively quickly.
These new simulations revise the current picture
of superhorizon-sized void evolution after first-order inflation.

\end{quotation}
\vspace{.5 in}
\centerline{{\it Submitted to Phys.~Rev.~D}}
\end{titlepage}
\newpage
\baselineskip=12pt

\centerline{\bf{I.  Introduction}}
\setcounter{section}{1}
\setcounter{equation}{0}
\renewcommand{\thefootnote}{\alph{footnote}}

The Big Bang model predicts the evolution of a homogeneous and
isotropic universe.  The confirmation of its predictions
(e.g.~the 3$K$ Planck spectrum of the microwave background,
the primordial abundances of $^4He$, $^3He$, $D$,
and $^7Li$ from Big Bang Nucleosynthesis, and the
number of light neutrino species) is stunning.
One is led
to question ${\it why}$ the universe can be described so well by a
homogeneous and isotropic model.

Inflation can provide the answer to this question.
It occurs when a scalar
field $\sigma$ has non-zero potential energy $V(\sigma)$ which dominates
the energy density of the universe.$^{{\ref{Guth}},{\ref{TEU}}}$
The key ingredients to all models
are that the universe expands superluminally during inflation and that
there is massive entropy generation afterward.
If the universe increases at least $10^{27}$ times
its original size, the flatness and horizon problems are solved;
the universe is dynamically
driven to homogeneity and isotropy.
The most interesting class of models is
first-order inflation.  Here,
the scalar field is trapped in the false
vacuum state of a strongly first-order potential.
Bubbles of true vacuum are nucleated at different spacetime points, and
the end of inflation occurs when the universe is filled with true-vacuum
bubbles of varying sizes.
The original model, Guth's ``old inflation'',$^{\ref{Guth}}$
does not work because it fails to percolate (fill) the universe with
true-vacuum bubbles.  More recent
models of first-order inflation which modify the gravitational or particle
sector
(e.g.~``extended inflation'' $^{\ref{LASTEIN}}$) are
promising as early-universe inflationary scenarios.
Because the universe expands as a power-law in time rather than exponentially,
percolation is guaranteed to occur.

The end of inflation occurs when true-vacuum bubbles of different sizes
fill all of space.
A confusing mess of scalar field dynamics then occurs as the bubbles
collide.$^{\ref{WatWid}}$
Because all the energy is contained in the bubble walls,
reheat occurs when
the $\sigma$-field gradient energy is converted into locally thermal
radiation.
After reheat, the standard
homogeneous and isotropic Big Bang model describes the evolution of that part
of the universe that had contained horizon-sized bubbles, since these would
have
been thermalized during reheat.
The very large superhorizon-sized bubbles (which were nucleated early on
during inflation), however, have traditionally
been a problem.$^{\ref{GoldZag}}$
After the $\sigma$-field in the bubble wall decays to relativistic particles,
a nearly empty void is formed.
Since these voids are much larger than the Hubble radius outside the void,
it has been thought that the inside
of the largest superhorizon-sized voids would remain empty until after
recombination, thus
producing unobserved temperature fluctuations of
order unity in the microwave
background.$^{{\ref{Wein}},{\ref{BLS}},{\ref{TurWeiWid}}}$
The longevity of these voids follows from assuming that a
superhorizon-sized
void expands conformally with spacetime.$^{{\ref{Wein}},{\ref{TurWeiWid}}}$
The earliest time at which thermalization could occur would then be when
the Hubble radius outside the void is of order the size of the void, since this
is the expected
$1^{\rm st}$-crossing time (i.e. the time for photons originally in the void
wall
to reach the origin).
If the void has comoving
size $r_0$, this occurs in time
$\Delta t=t-\ti\simeq H^{-1}_{\rm out}(\ti) (c^{-1}\RW/H^{-1}_{\rm
out}(\ti))^2$,
where the initial void radius is $\RW\equiv r_0 a(\ti)$,
$cH^{-1}_{\rm out}(\ti)$ is the Hubble radius outside the void,
$a(t)$ is the cosmic scale factor
and $\ti$ is the cosmic time after reheating.
Other authors suggest
that the void would fill in with radiation, but this is estimated
to occur on similar time scales.$^{\ref{LiddWan}}$
Because of this ``big bubble problem'', first-order inflation models
are fine-tuned in order to keep the production of large
bubbles at a minimum.  $^{{\ref{Wein}},{\ref{TurWeiWid}},{\ref{LiddWan}}}$

Motivated by the first-order (e.g. extended) inflation ``big-bubble problem'',
we present numerical studies of the evolution of a superhorizon-sized
general relativistic void embedded in a Friedmann-Robertson-Walker universe.
Although pressureless and thin-shell superhorizon-sized
voids have been studied in the
past,$^{{\ref{TolBon}},{\ref{ThinShell}}}$
this is the first study of general relativistic voids with
pressure and of arbitrary size and void wall structure.
We emphasize that our results are not
dependent on any particular inflation model.
These new simulations show that
opposite sides of a superhorizon-sized relativistic void interact
in a {\it very} short cosmic time, thereby suggesting that
these voids can also thermalize and homogenize on short time scales.

In Section II, we discuss the general relativistic spherically symmetric
metric in Lagrangian gauge and synchronous coordinates,
and present the equations to be solved numerically.
In Section III, the initial conditions,
boundary conditions and numerical techniques
used are described.
In addition, remarks about the deceleration of a void wall are made.
In Section IV, we derive the surprising result that the $1^{\rm st}$-crossing
time
of a superhorizon-sized general relativistic
void can be vanishingly short.
Section V contains numerical solutions for pressureless dust and
comparisons to exact Tolman-Bondi solutions.
A test of the code for the relativistic
Friedmann-Robertson-Walker solution is presented
in Section VI.
The numerical evolution of nonrelativistic, special relativistic, and
general relativistic voids with pressure
is examined in Section VII.
Here it is found that a relativistic superhorizon-sized void
collapses in the form of a strong shock moving at the speed of
light.  Thus, the collapse time is approximately the $1^{\rm st}$-crossing
time.
Finally, Section VIII contains a discussion of the results.

\centerline{\bf{II. Spherically Symmetric General Relativistic Fluids}}
\setcounter{section}{2}
\setcounter{equation}{0}

\centerline{\bf{A. The (Lagrangian) Metric}}

The most general spherically symmetric metric is$^{\ref{LLfluids}}$
\be
\label{eq:mostgen}
ds^2=c^2 l(t,r)dt^2+a(t,r)drdt+h(t,r)dr^2+
k(t,r)d\Omega^2,
\ee
where $d\Omega^2=d\theta^2+\sin^2\theta d\psi^2$ and $c$ is the speed of
light.
To this metric we can apply general transformations of the type
$t=f_1(t',r')$ and $r=f_2(t',r')$ without altering the spherical symmetry.
If we perform the necessary transformations to eliminate the
$~drdt$ term,
then Eq.~(${\ref{eq:mostgen}}$) can be written as
\be
\label{eq:metric}
ds^2=-c^2\Phi^2(t,r)dt^2+\Lam^2(t,r)dr^2+R^2(t,r) d\Omega^2,
\ee
where we have dropped the primes, and
where we require our coordinates
to be comoving with the fluid.
At time $\t$, the metric function $R(t,r)$
is the Eulerian distance that a fluid shell
labeled by $r$ is located from the center of coordinates.
More precisely, ~$2\pi R(t,r)$ is the spacelike circumference
of a sphere centered on the origin which contains all particles with comoving
coordinate $r$.
We have thus chosen the Lagrangian gauge with synchronous coordinates (Gaussian
normal coordinates).
Transformations of the form $\widetilde t=f(t)$ and $\widetilde r=g(r)$ can
still be made.
This metric was first used to study the
general relativistic collapse of stars to black
holes or neutron stars during supernovae.$^{{\ref{MisSharp}},{\ref{MayWhite}}}$

It is very important in numerical general relativity to carefully choose
the appropriate gauge and coordinates to best match the physical problem
to be studied,
We have thus specifically chosen the Lagrangian gauge and synchronous
coordinates to evolve a general relativistic
void embedded in an expanding Friedmann-Robertson-Walker (FRW) universe.
In the Lagrangian gauge, we gain maximal coverage
of the fluid in a numerical scheme.  This is important since the void is
embedded
in an expanding universe, so that we would continuously lose
mass shells in a Eulerian scheme.  In addition, the final results
are much more easily relatable to our own approximately FRW homogeneous and
isotropic universe.
We note that
asynchronous coordinates could instead be used
with the Lagrangian gauge.  However, here time and space
are mixed up so that the region outside a void is no longer spatially
homogeneous---the necessary initial conditions are not obvious and would need
to be determined using comoving synchronous coordinates
to initially set up and evolve the void.
Asynchronous slicing of space-time (e.g. polar slicing$^{\ref{PirBlud}}$)
is usually used numerically to study the collapse of a star
to a black hole in a flat non-expanding universe.  These
coordinates are necessary to study the
physically interesting mass zones outside the apparent horizon after a mass
shell
has crossed this horizon.  (In synchronous coordinates, once a mass shell
crosses
the apparent horizon, numerical integration stops so that the evolution of
the mass shells outside this horizon cannot be studied).
Since we are evolving an {\it underdense} region
here and do not expect apparent horizons to form, we do not need to resort to
these
coordinates.

Because we wish to embed a void in a FRW universe, we now relate the metric
functions from Eqn({\ref{eq:metric}}) to those from the familiar FRW metric:
\be
\label{eq:FRW}
ds^2=-c^2dt^2+a(t)^2\left(\frac{dr^2}{1-kr^2/c^2}+r^2 d\Omega^2\right),
\ee
where $a(t)$ is the cosmic scale factor, and $k$ is $-1$, $0$ or $1$ for
negative, zero and positive spatial curvature, respectively.
The solutions describe a universe which is homogeneous and isotropic
on each time slice.$^{\ref{TEU}}$
For the FRW metric then,
$\Phi=1$, $\Lam=a(t)/\sqrt{1-kc^{-2}r^2}$ and the Eulerian distance is
$R(t)=ra(t)$.

In this paper, the particles are assumed to be
everywhere in local thermal equilibrium, so that we can describe them
as a fluid.
The stress-energy tensor for a viscous fluid with energy density $\rho$
and pressure $p$
(measured in the frame of the fluid)
is \footnote{We set $c=1$ for the rest of this subsection}
\be
\label{eq:stressE}
T^{\alp\bet}=\rho u^\alp u^\bet+p P^{\alp\bet}-2\eta\sigma^{\alp
\bet}-3\zeta\Theta P^{\alp\bet}
\ee
where $u^{\alpha}$ is the fluid $4-$velocity, $P^{\alpha\beta}=
u^{\alpha}u^{\beta}+g^{\alpha\beta}$ is the projection operator,
$\sigma^{\alp\bet}=
1/2~(\der_\mu u^\alp P^{\mu\bet}+\der_\mu u^\bet P^{\mu\alp})-1/3~\Theta P^
\albe$ is the shear viscosity tensor, $\Theta=\der_\mu u^\mu$ is the fluid
expansion coefficient, and $\eta$ and $\zeta$ are arbitrary functions of
$r$ and $t$.  If $\eta=\zeta=0$, the fluid is non-viscous.
For Eqn(${\ref{eq:metric}}$) with comoving fluid 4-velocity
$u^{\alp}=(-\Phi^{-1},0,0,0)$,
${\sigma_r}^r=\beta$ and ${\sigma_\theta}^\theta=
{\sigma_\psi}^\psi=-1/2~\beta$, where
$\beta\equiv 2/3~(\dLam/\Lam-U/R)$, and
$\Theta=\dLam/\Lam+2U/R$.  (For the FRW
metric,
$\bet=0$ and $\Theta=3\dot{a}/a$).
The stress tensor is diagonal, with components
${T_t}^t=-\rho$, ~${T_r}^r=(p-\zeta\Theta)-2\eta\bet $,
and ${T_\theta}^\theta={T_\psi}^\psi$
$=(p-\zeta\Theta)+\eta\bet $.
We will employ a scalar artificial viscosity, $Q$, so that $\eta=0$ and
$\zeta\Theta= -Q$, where $Q$ will be significantly non-zero only
in areas of steep ``velocity'' gradients, as in a shock.
This is essential for stabilizing numerical shocks, as is
well known.$^{\ref{MortRich}}$
This viscosity will
dissipate enough energy on small scales so that the numerical solution
approaches
the exact solution in the limit that the grid spacing approaches zero.

\centerline{\bf{B. Fluids composed of massless particles}}

If a fluid consists of (effectively) massless, locally thermalized particles,
we
can relate $p$ to $\rho$ through the equation of state $p=p(\rho)$.
For photons (or particles with mass $\mu$ having local temperatures $T\gg\mu$),
$p=\rho/3$. $^{{\ref{TEU}},{\ref{Temp}}}$
Setting ${{\cal{R}}_\alp}^\bet-{g_\alp}^\bet {\cal{R}}=8\pi G_N {T_\alp}^\bet$,
where ${{\cal{R}}_\alp}^\bet$ is the Ricci tensor and $G_N$ is Newton's
constant,
and using the conservation equations $\der_\mu T^{\mu\alp}=0$, five independent
equations are found: ${G_0}^0={T_0}^0,~{G_1}^1={T_1}^1,~{G_0}^1=
{T_0}^1,~{T^{0\mu}}_{;\mu}=0,~{\rm and}~{T^{1\mu}}_{;\mu}=0$.
We define
\ba
\label{eq:Gamlam}
\Gamma&\equiv& R'/\Lambda,\\
\label{eq:Mprime}
M'&\equiv&4\pi c^{-2} \rho R^2R'
\ea
and $U\equiv \Phi^{-1}(\partial R/\partial t)$,
where the prime denotes differentiation with respect
to $r$.  The general relativistic equations can be written
\ba
\label{eq:nomasseq}
\dR &=&\Phi U \\
\label{eq:dUmss},
\dU &=&-\Phi\left(\frac{G_NM}{R^2}+\frac{4\pi G_N(p+Q)R}{c^2}\right)-
\frac{c^2\Gamma^2\Phi(p+Q)'}{(\rho+p+Q)R'} \\
\label{eq:dM}
\dM&=&-4\pi(p+Q)R^2\Phi U/c^2\\
\label{eq:drho}
\drho&=&-\Phi(\rho+p+Q)\frac{(R^2U)'}{R^2R'}\\
\label{eq:Phip}
\Phi'&=&-\Phi\frac{(p+Q)'}{\rho+p+Q} \\
\label{eq:defGamma}
\Gamma^2&\equiv& 1+(U/c)^2-2G_NM/(Rc^2),
\ea
where the dot denotes differentiation with respect
to $t$, and
where we have included an alternate definition for $\Gamma$.
There is also the auxiliary equation: $\dLam=\Lambda\Phi U'/R$.

The quantity $U\equiv \dot{R}/\Phi$ describes a particle's
``velocity'' as measured
in the frame of the fluid, since for $dr=d\theta=d\psi=0$,
the infinitesimal proper time that each observer measures is
$d\tau=\sqrt{-ds^2}
=c\Phi dt$.
For $G_N=0$, if a particle has velocity $v$, then
$\Gamma=1/\sqrt{1-(v/c)^2}$ and
$U=\Gamma v$ (see Eqn~({\ref{eq:Gaminterp}})); $\Gamma$ and $U$ represent the
two non-trivial
components of the 4-velocity of the fluid.  If $\Gamma\gg 1$, then
$U/c\simeq \Gamma$ and the fluid
is moving at relativistic velocities relative to a stationary observer.

We can rewrite Eqn~({\ref{eq:Mprime}}) in terms
of the proper volume element $d^3V=4\pi R^2\Lambda dr=4\pi R^2 R'dr/\Gamma$.
The ``mass-energy'' function $M$ then becomes
\be
\label{eq:Minv}
M(\t,\r)=c^{-2}\int d^3V ~\Gamma\rho.
\ee
For $G_N=0$,
because the volume along the radial direction is Lorentz-contracted
by the factor ~$1/\sqrt{1-(v/c)^2}=\Gamma$
and $\rho$ is the energy density measured in the frame of the fluid,
$\Gamma\rho$ is just the energy density of a fluid parcel
as measured by a stationary observer.
Therefore, $M(\r,\t)$ is just the total ``mass-energy''
contained within comoving coordinate $r$ at time $t$.

To better understand these equations, we relate them to the
FRW equations.
Recall the FRW results following Eqn~({\ref{eq:FRW}}):
$\Phi=1$, $R=ra$ and $\Lambda=a/\sqrt{1-kr^2/c^2}$.  From
Eqn({\ref{eq:nomasseq}}),
the ``velocity'' is $U=\dot{R}=r\dot{a}$.  Since $R'=a(t)$,
$\Gamma=\sqrt{1-kr^2/c^{2}}$ from Eqn({\ref{eq:Gamlam}}).
For a spatially flat ($k=0$)
FRW model then, $\Gamma=1$ even though the
fluid at $R$ is moving away from
the origin with velocity $U$.  Thus in general relativity,
$\Gamma$ in not the relativistic gamma-factor of the fluid with respect
to the origin.
Using the fact that
$M=4\pi c^{-2} \rho R^3/3$, we see that Eqn~({\ref{eq:defGamma}}) is just
Friedmann's equation, $H^2=({\dot{a}}/a)^2=8\pi G_Nc^{-2}\rho/3-kc^2/a^2$,
with $H\equiv \Phi^{-1}\dot{a}/a=U/R$.

In a spatially flat FRW universe then,
$U=c^{-1}R\sqrt{8\pi G_N \rho/3}=\sqrt{2G_NM/R}$.
If we set $U\equiv U_{\rm GRAV}+U_{\rm PEC}$, where
$U_{\rm GRAV}\equiv 2G_NM/R$ is the gravitational velocity
and $U_{\rm PEC}$ is the peculiar
velocity, $\Gamma$ can be
expressed in terms of these velocities:
$\Gamma^2=1+(U_{\rm PEC}/c)^2+2U_{\rm PEC}U_{\rm GRAV}/c^2$.

For a non-viscous fluid with $p=(\gamma-1)\rho$,
Eqn({\ref{eq:Phip}}) (the conservation of momentum equation,
${{T^{\mu}}_{1}}_{;\mu}=0$)
can be integrated exactly
to give $\Phi(t,r_2)=\Phi(t,r_1)[\rho(t,r_1)/$
$\rho(t,r_2)]^{(\gamma-1)/\gamma}$.
We are interested in evolving a void embedded in a FRW homogeneous
and isotropic universe, so that
$p'=\rho'=0$ outside the void.
We define $\Phi_{\rm out}$ and $\rho_{\rm out}$ to be the spatially
constant values
of $\Phi$ and $\rho$ outside the void.  (In what follows, the subscripts
``in'' and ``out'' represent the spatially constant values inside and
outside the evolving
void region, respectively).  Taking $\gamma=4/3$, we find
\be
\label{eq:Phirel}
\Phi=\Phi_{\rm out}\left(\frac{\rho_{\rm out}}{\rho}\right)^{1/4}.
\ee
The fact that this equation can be solved exactly is
important for calculating the $1^{\rm st}$-crossing time
for general relativistic voids, as will be discussed in Section IV.

To gain some physical insight into $\Phi$, we calculate the potential energy
of a fluid distribution.
A comoving observer has $dr=d\theta=d\psi=0$, so that the proper
acceleration measured by this observer is $a_r=\Phi^{-1}\dU$.  Since the
force is radial, we can write $a_r=-\partial\phi/\partial R$, where $\phi$
is the potential.  Using Eqn({\ref{eq:dUmss}}) and integrating,
we can write the potential as the following sum:
$\phi(t,R)=\phi_{\rm GRAV}(t,R)+\phi_{\rm FLUID}(t,R)$, where
\ba
\phi_{\rm GRAV}&=&4\pi G_N \int_0^R\left[{M}/{R^2}+c^{-2}p R\right]dR\\
\phi_{\rm FLUID}&=& \int_0^R  c^{2}\Gamma^2 {dp}~/~({\rho+p})
\ea
for $Q=0$.  For a void with $p=\rho/3$, we obtain the usual gravitational
potential
$\phi_{\rm GRAV}\simeq 4\pi G_N\rho_{\rm in}R^2/3=G_N M/R$ inside the void,
and $\phi_{\rm GRAV}\simeq 4\pi G_N\rho_{\rm out}R^2/3\simeq G_N M/R$ outside.
If in addition we choose $\Gamma(\ti,R)=1$ initially, then
$\phi_{\rm FLUID}(\ti)=\ln[~\rho(\ti,R)/\rho_{\rm in}(\ti)~]^{1/4}=
-\ln[~\Phi(\ti,R)/\Phi_{\rm in}(\ti)~]$.
(Note that the contribution to the fluid potential is zero inside,
but (potentially much) greater than zero outside the void.
Thus it can substantially increase the already large potential outside
the void).
Therefore, $\Phi$ is proportional to the exponential of the fluid potential,
$\phi_{\rm FLUID}$.

\centerline{\bf{C. Fluids composed of massive particles}}

Suppose instead we consider a fluid which consists of particles
of mass $\mu$ and with arbitrary temperature $T$.  Then, the total
energy density of the fluid
is the mass energy density plus
the internal energy density.
Denoting the (proper) mass
density ($\mu$ times the number density) by $n(\t,r)$ and
the internal energy per unit mass by $\ep(\t,r)$,
\be
\label{eq:defrho}
\rho= c^2 n(1+\ep/c^2).
\ee
Here we have traded one unknown for two because in
general the pressure depends not only on $\rho$ but also on $n$.
For a fluid composed of relativistic (nonrelativistic) particles, $\ep/c^2> 1$
($\ep/c^2< 1$).
If the fluid obeys the perfect gas law, then its pressure is
\be
\label{eq:defp}
p=(\gam-1) n\ep.
\ee
For a highly relativistic species with $\gam=4/3$, the energy
density is three times the pressure: $\rho= n\ep=p/(\gam-1)=3p$, whereas for
a highly nonrelativistic fluid, the
energy density is much larger than the
pressure: $\rho\simeq nc^2=p/[(\gam-1)\ep/c^2]\gg p$.
We assume that the total number of particles per comoving volume is constant:
$\der_\mu(nu^\mu)=0$. $^{\ref{Schutz}}$
Using Eqn({\ref{eq:Gamlam}}), this can be integrated to give
\ba
\label{eq:defn}
f(r) &=& 4\pi nR^2R'/\Gamma \\
\label{eq:deff}
&\equiv&r^2
\ea
where $f(r)$ is an arbitrary function depending only
on the coordinate $r$.
Specifying $f(r)$ completely
fixes the arbitrariness of the metric functions
under transformations in $r$ (as discussed after Eqn~({\ref{eq:metric}})).
This particular definition for $f$ is necessary in order to write the
difference
schemes in a geometrical way that allows shocks and explosions to be
numerically stable at the origin.$^{\ref{MortRich}}$

We can now rewrite the full set of general relativistic equations
(({\ref{eq:nomasseq}})-({\ref{eq:defGamma}}))
as$^{{\ref{MisSharp}},{\ref{MayWhite}}}$
\ba
\label{eq:eom}
\dR&=&\Phi U \\
\label{eq:dU}
\dU&=&-\Phi\left(\frac{G_NM}{R^2}+\frac{4\pi G_N(p+Q)R}{c^2}\right)-
\frac{4\pi\Gamma\Phi R^2(p+Q)'}{w r^2} \\
\dM&=&-4\pi(p+Q)R^2\Phi U/c^2\\
\label{eq:intn}
\dn&=&-\frac{n\Phi(R^2U)'}{R^2R'}\\
\label{eq:dep}
\dep&=&-\frac{4\pi\Phi (p+Q)(R^2U)'}{\Gamma~r^2} \\
\label{eq:defPhi}
\Phi'&=&-\Phi\frac{(p+Q)'}{nwc^2},
\ea
where $\Gamma$ is given by Eqn({\ref{eq:defGamma}}) and
$w \equiv 1+(\ep+p/n)/c^2$ is the relativistic enthalpy.
Equations ({\ref{eq:eom}})-({\ref{eq:defPhi}}) (along with the
definitions for $\Gamma$ and $w$ given in the previous sentence)
are the set used in the numerical code.\footnote{For technical difficulties
in the general relativistic case,
we determine $M$ via the $\dM$ equation rather than the
$M'$ equation.  In addition, we determine $n$ via the
$\dn$ equation rather than through the analytical solution
$n=\Gamma r^2/(R^2 R')$, because too much intrinsic
viscosity is introduced for special relativistic voids otherwise.}

When the kinetic
energy of each particle is much less than its mass energy $\ep/c^2 \ll 1$
(or $T/\mu\ll 1$),
we obtain the nonrelativistic Lagrangian fluid equations.$^{\ref{CourFrei}}$
(They can also be obtained by setting
$c^2\rightarrow \infty$ in Eqns~({\ref{eq:eom}})-({\ref{eq:defPhi}})).
For future reference, in this limit $\Phi\rightarrow 1$, $\Gamma \rightarrow
1$,
and $w\rightarrow 1$
so that $U=\dot{R}$ is the fluid velocity,
$M(t,r)=r^3/3$ is the total mass within $r$ and
$n=r^2/(R^2R')$ is the mass density.

The artificial viscosity used here is given by
Equation ({\ref{eq:artvis}}), and is generalized from the expression used by
previous workers:$^{{\ref{VonNRich}},{\ref{MayWhite}}}$
\ba
\label{eq:Visc}
Q&=&k^2n(1+\ep/(\Gamma c^2))(U')^2dr^2/~\Gamma~~~{\rm for~U'<0}\nonumber\\
Q&=&0~~~~~~~~~~~~~~~~~~~~~~~~~~~~~~~~~~~~~{\rm otherwise}.
\ea

Consider the behavior of $n$ and $p$ in the limit that the entropy, $S(t,r)$,
within $r$ is conserved.
Using the thermodynamic relation  $TdS=d(\rho/n)+pd(1/n)$, we find
$\dep=p\dn/n^2$.  Then, the pressure and internal energy for a shell
labeled by $r$ are related to their initial values by
$p\propto n^{\gamma}$ and $\ep\propto p^{(\gamma-1)/\gamma}$.
In the ultrarelativistic limit with $\gamma=4/3$,
$n \propto\rho^{3/4}$ and $\ep\propto\rho^{1/4}$.
In addition, if the fluid is a ploytrope (i.e. isentropic), $S'=0$.
Using $TdS=d(\rho/n)+pd(1/n)$ again, we find the
familiar result$^{\ref{LLcfluids}}$
\be
\label{eq:isentrop}
\frac{\ep(t,r_1)}{\ep(t,r_2)}= \left(\frac{n(t,r_1)}{n(t,r_2)}\right)^
{\gamma-1}=\left(\frac{p(t,r_1)}{p(t,r_2)}\right)^{(\gamma-1)/\gamma}.
\ee
For ultrarelativistic fluids, $\rho=n\ep$ and therefore
$\ep\propto\rho^{(\gamma-1)/\gamma}$ and $n\propto\rho^{1/\gamma}$.
If $\gamma=4/3$, $\ep\propto \rho^{1/4}$
and $n\propto\rho^{3/4}$.
This is the property of a relativistic fluid, since
then $\rho\propto T^4$ and $n\propto T^3$.

\centerline{\bf{D. Fluid Deceleration in the Void Wall}}

In this section we show that if the outward peculiar velocity of the wall of
a general relativistic void
is very large, then the deceleration of this wall can be enormous.
This ``damping force'' is responsible for slowing down and collapsing
a superhorizon-sized
void.  Without this fluid force, the void would expand (not collapse) from
gravitational forces(see V).

In section II-B, we saw that for special relativistic fluids ($G_N=0$),
$U$ is the net outward momentum per particle mass $\mu$.
Similarly, if in the general relativistic case $U^2\gg U_{\rm GRAV}^2=2G_NM/R$,
the gravitational attraction inward is negligible
and the dynamics will be dominated by special relativistic
effects;  $U$ can be loosely interpreted as the peculiar momentum per particle
per mass.
If $\Gamma\gg 1$ and $U>0$
in the wall of a void, the wall moves outward with
momentum much greater than the gravitational attraction inward.
We now investigate what happens to this wall.  From Eqn~({\ref{eq:dU}}),
the ``conservation of momentum'' equation is
\ba
\label{eq:totconsmom}
nw\Phi^{-1}\dU&=&-nw\left(\frac{G_NM}{R^2}+\frac{4\pi G_N(p+Q)R}{c^2}\right)-
\Gamma^2\frac{(p+Q)'}{R'}.
\ea
The functions $\Gamma$, $\Phi$, $n$, $p$, $Q$, $M$, $w$ and $R'$ are always
positive.  If $U>0$, then at the inner edge of the void wall where $(p+Q)'>0$,
$\dU$ will be negative---the fluid there
is decelerated.
This deceleration is due to both gravitational and fluid forces.  We
examine the fluid force contribution only.
Using Eqns~({\ref{eq:defGamma}}), ({\ref{eq:Mprime}}) and ({\ref{eq:dU}})
we find that
\be
\label{eq:dGexa}
\dG=-\Gamma U\Phi \frac{(p+Q)'}{nwc^2R'}.
\ee
Again, if $(p+Q)'>0$ and $U>0$, $\dG$ will be negative and
the fluid particles there lose their energy per particle mass.

Suppose the wall has a very large outward momentum so that
$U/c\gg\sqrt{2G_NM/(Rc^2)}>1$.  Then $\Gamma\simeq U/c$.
Setting $p=\rho/3$ and $Q=0$ as for a relativistic, non-viscous
fluid,
\be
\label{eq:Gammadamp}
\dG=-\Gamma^2 \frac{\Phi\rho'}{4\rho R'}\simeq -\Gamma^2~
\frac{\rho_{\rm out}^{1/4}~\rho'}{4~\rho^{5/4} ~R'}~,
\ee
where we have used the solution for $\Phi$ from Eqn~({\ref{eq:Phirel}}).
We consider the deceleration of fluid shells at the inner edge of a steep void
wall.
We can write $\rho'/R'\simeq \Delta\rho_{\rm wall}/\rwt$,
where $\Delta\rho_{\rm wall}$ is the difference in the energy density
over the width of the wall, and $\rwt$ is
the thickness of the
wall.  Then, since $\Delta\rho_{\rm wall}\simeq \rho_{\rm max}$,
where $\rho_{\rm max}$ is the maximum wall energy density,
we can approximate $\dG$ by
\be
\label{eq:dGapp}
\dG\simeq -\frac{\Gamma^2}{4}\left(\frac{\rho_{\rm out}\rho_{\rm max}^4}
{\rho^5}\right)^{1/4} \rwt^{-1}.
\ee
Initially, except for the first factor of $\Gamma^2$,
all factors on the right hand side of the previous equation
are independent of $\Gamma$.
Thus, $\dG\propto -\Gamma^2$, which can be a very large damping
factor!
The second factor is proportional to
$\rho_{\rm out}\rho_{\rm max}^4/\rho^5\leq (\rho_{\rm out}/\rho)^5$,
so for the mass shells on the innermost part of the wall (i.e. those
shells with the
smallest values of $\rho/\rho_{\rm out}$) this factor can be enormous.
The third factor is $\rwt^{-1}$, so the thinner the wall, the
faster it will slow down.
If the second and third factors change slowly enough with time,
then the slow-down time for the wall is roughly
independent of its initial value of $\Gamma_0$, since
$-\int_{\Gamma_0}^1 d\Gamma/\Gamma^2\simeq 1\ga (\rho_{\rm out}/\rho)^5
\rwt^{-1}\Delta t/4$ for $\Gamma_0\gg 1$.
This could be an extremely important result, and would imply that
the initial peculiar wall velocity could never be large enough to
cause a void to expand for an arbitrarily long time.
However, because the second and third factors in
Eqn~({\ref{eq:dGapp}}) will change with time and depend on
$\Gamma$, this is only a crude guess.

As a concluding remark, we note that the wall of a void formed during
first-order inflation has an enormous outward peculiar velocity.  This enormous
velocity has been thought to cause a void
to expand ``indefinitely''.  However, with such a large deceleration of the
void wall, it will slow down in a finite (and possibly small)
amount of time.
A future paper will explore this numerically.$^{\ref{Vasha}}$

\centerline{\bf{III. Initial Conditions, Boundary Conditions, and Numerical
Techniques}}
\setcounter{section}{3}
\setcounter{equation}{0}

\centerline{\bf{A. Initial Conditions}}

The grid used in this code is initially
equally spaced: $\dr\equiv R(\ti)_{j+1}-R(\ti)_j=constant$,
where the subscript $j$ denotes the spatial grid point number and ranges from
$j\in [0,j_B]$, where $j_B$ is its value at the outer boundary.
(Note from Eqns~({\ref{eq:defn}}) and ({\ref{eq:deff}}) that in general,
$\Delta r\neq constant$ then).
The $0^{th}$ and $1^{\rm st}$-grid points are located at $R_0=-\dr/2$ and
$R_1=\dr/2$
respectively.  Thus, $R_j(t_i)=R_1(t_i)+(j-1)\dr$.

The initial conditions for the functions at $\ti$
are determined as follows.  The viscosity is set to zero: $Q(\ti,R)=0$.
The energy density $\rho(\ti,R)$ (or the ``mass-energy'' $M(\ti,R)$), the
``velocity'' $U(\ti,R)$ (or $\Gamma (\t_i,R)$)
and the specific internal energy, $\ep(\ti,R)$, are chosen as functions
of the radius $R(\ti)$.
(If $M$ is specified initially instead of $\rho$, we determine $\rho$
via Eqn.~({\ref{eq:Mprime}})).
For the cases run in this paper however,
we will initially specify the fluid to be a polytrope (constant entropy
on the initial time slice (II-C))
and $\ep/c^2\gg 1$ or $\ep/c^2\ll 1$.  Using the relations found at the
end of Section II-C,
$\ep(\ti,R)$ is determined once the internal energy
$\ep(\ti,\Rmx)$ is specified at the outer boundary ($\Rmx\equiv R(\ti)_{j_B}$):
\ba
\ep(\ti,R)&=&\ep(\ti,\Rmx)\left[~{\rho(\ti,R)}/{\rho(\ti,\Rmx)}~\right]^
{~(\gamma-1)/\gamma}~~
{\rm for}~\ep/c^2\gg 1 \nonumber\\\
\ep(\ti,R)&=&\ep(\ti,\Rmx)\left[~{\rho(\ti,R)}/{\rho(\ti,\Rmx)}~\right]^
{~(\gamma-1)}~~~~{\rm for}~\ep/c^2\ll 1
\ea
We then determine $n$ and $p$ by
$n(\ti,R)=\rho(\ti,R)/(1+\ep(\ti,R))$ and $p(\ti,R)=(\gam-1)n(\ti,R)\ep(\ti,R)$
respectively.
Next, we find $r$ (and $M$ if it is not initially specified) by integrating
outward from $\r=0$ using the
$4^{th}$-order Runge-Kutta method:
\ba
r&=&\left(~3\int_0^{R(\ti,r)} {4\pi nR^2 dR}/\Gamma \right)^{1/3}\nonumber\\
& &\left(~{\rm and}~~~M(\ti,R)=\int_0^R n(1+\ep/c^2) R^2 dR~\right),
\ea
where we have omitted the $(\ti,R)$'s for clarity.
Finally, $\Phi$ is found by integrating Eqn~({\ref{eq:defPhi}})
inwards from the outer boundary
(again using the $4^{th}$-order Runge-Kutta method)
once $\Phi(\ti,\Rmx)$ is specified.

We are interested in evolving a non-expanding,
empty void embedded in a FRW universe.
We will therefore initially set the energy density outside the void
($\rho_{\rm out}\equiv \rho(\ti,\Rmx)$)
to be constant and equal to the spatially flat FRW value.
Then, $H_{\rm out}^2(\ti)\equiv H^2(\ti,\Rmx)=
(\p/\ti)^2=8\pi G_N c^{-2}\rho_{\rm out}(\ti)/3$,
where $a_{\rm out}(t)=a(\ti)(t/\ti)^\p$ is the cosmic scale factor
and $cH_{\rm out}^{-1}$ is the Hubble radius
outside the void.
(Note that for the $k=0$ FRW universe, $r=(4\pi n(\ti))^{1/3}R(\ti,r)$, or
$a(\ti)=(4\pi n(\ti))^{-1/3}$).
For $p=\rho/3$ and $p=0$, ~$\p=1/2$ and $\p=2/3$ respectively.
We can quantify the initial size of
a void by measuring its radius relative to
the horizon size outside the void initially:
 $c^{-1}\RW/H_{\rm out}^{-1}(\ti)=c^{-1}\p \RW/\ti$, where
$\RW$ is the void ``radius''.
Following past convention$^{\ref{TEU}}$, we loosely equate the Hubble radius
with the horizon in the phrase ``superhorizon-sized''.
(Horizon in this context
is not to be confused with the particle horizon.)
A void is defined to be superhorizon-sized if
$c^{-1}\RW/H_{\rm out}^{-1}(\ti)> 1$
and subhorizon-sized if $c^{-1}\RW/$ $H_{\rm out}^{-1}(\ti)< 1$.
In addition, $\Gamma_{\rm out}(\ti)=1$ (see II-B) (or $U=\sqrt{2G_NM/R}$ )
and $\Phi_{\rm out}(\ti)\equiv 1$ (see II-A).

In this paper, we consider voids which are initially either compensated or
uncompensated in energy density and which have the following
distributions.  The ``mass-energy'' function for compensated voids is defined
to be
\be
\label{eq:compen}
M(\ti,R)=.5\rho_{\rm out}(\ti)\left[(1+\tanh x)+\alpha(1-\tanh x)\right]R^3
(\ti)/3,
\ee
where $x\equiv \x$,
$\rw$ is the wall thickness and $\alpha$
is a specified constant less than or equal to $1$.
Because $M$ reaches its spatially flat FRW value outside the void, the energy
density missing from the void has been put in the wall.
For uncompensated voids, the energy density is instead initially
specified.  It is
\be
\label{eq:uncompen}
\rho(\ti,R)=.5\rho_{\rm out}(\ti)[(1+\tanh x)+\alpha(1-\tanh x)].
\ee
Here, the energy density missing from the void has not been put into the wall.
Therefore, the region outside a compensated void will always be a spatially
flat ($k=0$) FRW universe, whereas the region outside an uncompensated void is
a
negative spatially curved ($k<0$) FRW universe.

The inside of the void is chosen to be homogeneous initially.
Since we want it to be nonexpanding in the
limit that it is empty ($\rho\rightarrow 0$),
we choose the inside of the void to be a spatially flat ($k=0$)
FRW ``mini'' universe.
We reason as follows.
Friedmann's equation inside the void is
$H^2_{\rm in}\simeq 8\pi G_N c^{-2}\rho_{\rm in}/3-kc^{-2}r^2/R^2$ (see II-B),
where the subscript `${\rm in}$' denotes quantities inside the void,
$R=ra_{\rm in}$ and
$H_{\rm in}\equiv U_{\rm in}/R=\Phi_{\rm in}^{-1}\dot{a}_{\rm in}/a_{\rm in}$.
In the limit that $\rho\rightarrow 0$, the inside of the void is
non-expanding ($\dot{a}_{\rm in}=0$) only if $k=0$.
Since we cannot numerically choose
$\rho_{\rm in}=0$ (because $\Phi_{\rm in}=\infty$ from Eqn({\ref{eq:Phirel}})),
we would like the inside of the void to
not expand on time scales that the outside region
expands in.   This is
satisfied if $\rho_{\rm in}/\rho_{\rm out}\ll 1$
because the Hubble time inside the void
is much larger than that outside the
void---~$H^{-1}_{\rm in}/H^{-1}_{\rm out}=\sqrt{\rho_{\rm out}/\rho_{\rm in}}$.
(As a check, Eqns.~({\ref{eq:dU}}) and ({\ref{eq:dep}}) show that as
$U\rightarrow 0$, $\ep\rightarrow 0$ and $p\rightarrow 0$, then
$\dU\rightarrow 0$ and $\dep\rightarrow 0 $).
Because the void is approximately homogeneous,
the ``mass-energy'' inside the void
is $M(\ti,R)\simeq c^{-2}\rho_{\rm in} R^3/3$. Finally, we set
$\Gamma_{\rm in}(\ti)=1$ inside the void since
$\Gamma=\sqrt{1-kr^2/c^{2}}$.  The ``velocity is then $U=\sqrt{2G_NM/R}\simeq
c^{-1}R\sqrt{8\pi G_N\rho/3}$.

We will only
consider two types of initial velocity profiles in this paper.  The first is
$U=U_{\rm GRAV}=\sqrt{2G_NM/R}$ (or $\Gamma(\ti,R)=1$).
This specifies that the outward
``velocity'' of each particle is just large enough to compensate for
the inward gravitational attraction (i.e. the peculiar velocity, $U_{\rm PEC}$,
is zero).  The second is
\be
\label{eq:velbig}
U(\ti,R)=c^{-1}R\sqrt{{8\pi G_N\rho}/{3}}.
\ee
For $R\ll \RW-\rw~{\rm and~}R\gg \RW+\rw$,
$\Gamma\simeq 1$.  In the wall region, however,
$\Gamma > 1$.  This corresponds to an initial net outward ``peculiar momentum''
of the particles in the wall.

\centerline{\bf{B. Boundary Conditions}}

We set the outer boundary conditions for all times to be those for a
homogeneous
fluid.
This specification works well in practice as long as the action
is taking place away from this boundary.
Thus we set $n'=\ep'=p'=0$ and $Q=0$ at $j=j_B$, where $j_B$ is the
grid point number for the outermost comoving coordinate.
In addition, we set $\Phi_{j_B}(\t)\equiv 1$, its FRW value
(see Eqn~({\ref{eq:FRW}}).
Note that specifying $\Phi_{j_B}(\t)$ completely eliminates
the arbitrariness of the time coordinate, as discussed
after Eqn~({\ref{eq:metric}}).
The present definition sets $t$ to be the FRW cosmic time outside
the void.  Thus if an initially inhomogeneous fluid becomes homogeneous,
then from Eqn({\ref{eq:defPhi}}),
$\Phi(t,r)=1$ everywhere, and $t=constant$ hypersurfaces
correspond to $t=constant$ FRW homogeneous and isotropic hypersurfaces.

It is possible to determine the outer boundary conditions for all $\t$
by solving the
equations with $Q=p'=0$.  One is then left with two $1^{st}-$order ordinary
differential
equations to solve.
The boundary conditions determined this way, however, give larger errors
than the ones shown below, and
therefore were not used.
We instead integrate
\ba
\dR&=&U \nonumber\\
\dU&=&-\left({G_NM}/{R^2}+{4\pi G_NpR}/{c^2}\right)\nonumber\\
\dM&=&-4\pi~p R^2U/{c^2}.
\ea
using the MacCormack method.
We then determine $n$, $\ep$ and $p$ from Eqns~({\ref{eq:isentrop}}) and
Eqn~({\ref{eq:defp}}) by setting
$n_{j_B}=n_{{j_B}-1}$,
$\ep_{j_B}=\ep_{j_B}(\ti)\left[~{n_{j_B}(t)}/{n_{j_B}(\ti)}~\right]^{\gamma-1}$
and $p_{j_B}=(\gamma-1)n_{j_B}\ep_{j_B}$.

Finally, reflecting boundary conditions are used at the inner boundary:
$R^n_0=-R^n_1$, $U^n_0=-U^n_1$, $p^n_0=p^n_1$, $n^n_0=n^n_1$,
$\ep^n_0=\ep^n_1$ and $Q^n_0=Q^n_1$,
where the superscript $n$ refers to values on the n$^{th}$ time slice.

\centerline{\bf{C. Numerical Integration}}

The $\dU,~\dep,~\dR$, $\dn$ and $\dM$ equations
are integrated using the 2-step
MacCormack predictor-corrector method.$^{\ref{BerHobSm}}$
In Appendix B, we give the exact form for the difference equations which
allows inbound shocks to rebound off the origin.
To illustrate the MacCormack method, we show the predictor
and corrector steps
for $\dn$ as an example.
We continue to use the convention that $n_j^i$ is the value of $n$ on the
$j^{th}$
spatial grid point and at the $i^{th}$ time step.
Suppose we know all quantities on the $i^{th}$ time slice.
We would like to determine them on the $(i+1)^{th}$ time slice.
First we predict the new quantities (with forward differencing)
using the functional values on the $i^{\rm th}$ time slice:
\be
{n_p}^{i+1}_{~j}=n^{i}_j-\Delta t~n^{i}_j \Phi^{i}_j~
\frac{{R^{i}_{j+1}}^2 ~U^{~i}_{j+1}-{R^{i}_j}^2 ~U^{~i}_{j}}
{{R^{i}_j}^2(R^{~i}_{j+1}-R^{~i}_{j})}.
\ee
After using similar methods to obtain ${U_p}^{i+1}_{~j}$, ${R_p}^{i+1}_{~j}$,
${M_p}^{i+1}_{~j}$ and ${\ep_p}^{i+1}_{~j}$ (and setting ${p_p}^{i+1}_{~j}=
(\gamma-1){n_p}^{i+1}_j {\ep_p}^{i+1}_j$) for all $j$,
we integrate the $\Phi'$ equation inwards from $j=j_B$
using the $4^{th}$-order Runge-Kutta method with linear interpolations
to determine ${\Phi_p}^{i+1}_{~j}$ for all $j$.
We integrate again, using the predicted values
obtained above (with backward differencing), and then average these
with the previously predicted values:
\be
{n}^{i+1}_{~j}=.5\left({n_p}^{i+1}_{~j}+ n^{i}_j-
\Delta t~{n_p}^{i}_j {\Phi_p}^{i}_j~
\frac{{{R_p}^{i}_{j}}^2 ~{U_p}^{~i}_{j}-{{R_p}^{i}_{j-1}}^2 ~{U_p}^{~i}_{j-1}}
{{{R_p}^{i}_j}^2({R_p}^{~i}_{j}-{R_p}^{~i}_{j-1})}\right).
\ee
We obtain the other values in a similar way, and then
integrate again to find $\Phi^{i+1}_{~j}$.

As is well known,$^{\ref{MortRich}}$ it is important to choose small
enough time steps $\Delta t$ to satisfy the Courant condition.
This condition requires $\Delta t$ to
be smaller than the time taken for sound to cross from
any one grid point to the next.
The speed of sound for relativistic fluids is
$c_S=\sqrt{(\partial p/\partial
\rho)_S}$.$^{{\ref{MisSharp}},{\ref{MayWhite}}}$
Using the fact that $TdS=d(\rho/n)+pd(1/n)$, we find
\be
\label{eq:sos}
c_S=\sqrt{\frac{\gamma p}{nw}}.
\ee
As $c^2\rightarrow\infty$, $w=1$ and we obtain
the usual expression for the speed of sound in nonrelativistic fluids.
Note that for $\ep/c^2\gg 1$,
$c_s=c\sqrt{\gamma-1}=.57c$ for perfect fluids with $\gamma=4/3$.
The Courant condition requires the proper time per proper distance to
be equal to $C/c_S$, where
$C$ is the Courant number and
is approximately $1/3$ to $1/2$ for strong shocks.
Since the infinitesimal proper time and proper distance are $\Phi\Delta t$
and $\Lambda \Delta r=\Delta R/\Gamma$ respectively, the Courant
condition becomes $\Delta t_i^n=C~(R_{i+1}^n-R_i^n)/
(\Gamma_i^n \Phi_i^n (c_S)_i^n)$. $^{\ref{MayWhite}}$
In addition, when $G_N\neq 0$, regions of the
universe expand and contract.
We require the time steps to be small enough
for the functions $n,~\ep,~{\rm and~}M$ to change
sufficiently slowly.  We therefore set
$\Delta t_i^n =\f~(l_i^n/\dl_i^n)$ to be the maximum time step
allowed for $l=n,~\ep~{\rm and~}M$, where
$\f$ is a constant less than one.

After the $n^{th}$ corrector step, $\Delta t_i^n$
is calculated for all $i$, and $\Delta t^{n+1}$ is set to the
smallest value obtained:
\be
(\Delta t)^{n+1} = {\rm min}\left(~\Delta t_{\rm max},~~C~
\frac{R_{i+1}^n-R_i^n}{\Gamma_i^n\Phi_i^n (c_S)_i^n}~, ~~
\left[\f~\frac{n_i^n}{\dn_i^n},~~\f~\frac{\ep_i^n}{\dep_i^n},~~
\f~\frac{M_i^n}{\dM_i^n}~\right]_{{\rm if~} G_N\neq 0}\right),
\ee
where $\Delta t_{\rm max}$ is a specified upper bound, if
desired.$^{\ref{local}}$

We apply a convergence test to the code for
test problems where analytic solutions are available.
The relative error in $q$, where $q$ denotes any quantity,
is defined to be
\be
\label{eq:ei}
e_i=|q_i-\widetilde{q}(r_i)|/\widetilde{q}(r_i),
\ee
where $\widetilde{q}(r_i)$ is the exact solution and $q_i$ is the numerical
solution.  We obtain a global measure of the error by defining
\be
\label{eq:conv}
L_1=\frac{1}{N}\sum_{i=1}^N e_i,
\ee
where $N$ is the total number of grid points.
This error is proportional to the grid spacing to some power:
$L_1\propto \Delta R^s$, where $s$ is the convergence rate.
If $s\simeq 2$, the code is second-order, as desired.
These tools have been used previously to test codes in other
applications.$^{\ref{Stoneetal}}$

\centerline{\bf{IV. ${\bf 1^{\rm\bf st}}$-Crossing Time for Relativistic
Voids}}
\setcounter{section}{4}
\setcounter{equation}{0}

In this section, we first review the standard lore for the evolution of
superhorizon-sized voids, and then calculate the $1^{\rm st}$-crossing time
(the cosmic time taken for a photon initially at the inner edge of the void
wall to reach the origin) in this picture.
We then calculate the {\it actual} $1^{\rm st}$-crossing time.
If at the $1^{\rm st}$-crossing time
the fluid were approximately homogeneous and isotropic,
distortions from the original void
would be negligible and a (nearly)
FRW universe would result everywhere.

It has been suggested that superhorizon-sized voids formed from first-order
inflation would conformally expand with spacetime during the
radiation-dominated period.$^{{\ref{Wein}},{\ref{TurWeiWid}}}$
Thus, the size of a void at time $t$ would be $R=r_0a(t)$, where $r_0$ is
the comoving coordinate of the void and $a(t)$ is the cosmic scale factor.
There are several justifications to support this belief.
First, small density perturbations conformally expand with spacetime.  Second,
vacuum bubbles conformally expand during inflation after an initial
short growing period.  Third, the time taken for the void wall
to slow down is expected to be enormous because the initial
outward momentum of the void wall is enormous.
However, this reasoning is not enough to conclude that a superhorizon-sized
void conformally expands with spacetime.
First, a void is not a small perturbation in spacetime.
Thus linear results can not be applied to the description
of a void.  Second, it is the negative pressure that causes
vacuum bubbles to expand; a similar configuration having positive pressure
would
instead acclerate inward.\footnote{In the bubble
wall, $G_N=0$, $|p|'>0$ and $p<0$. $^{\ref{Guth}}$  Using Eqn({\ref{eq:dU}}),
we see that the acceleration is positive, so that the bubble wall moves outward
during inflation.  But for normal positive pressure with $p'>0$, the
acceleration
is {\it negative}, so that the wall must accelerate inward.}
Third, although part of the wall may still move out,
the deceleration of the inner void wall is extremely large
(see II-D), so that the void can still
collapse (see VII).

If a superhorizon-sized void were to conformally expand in spacetime, then
the earliest time at which thermalization and homogenization can occur is when
the horizon is of order the size of the void, since this is the expected
$1^{\rm st}$-crossing time.  If the void has comoving
size $r_0$ and the outside Hubble radius (that outside the void)
is $c\Hm$, this occurs when
$r_0a(t)=H^{-1}_{\rm out}(t)$.  For evolution during the radiation-dominated
epoch, $H_{\rm out}^{-1}(t)=2t$ and $p=\rho/3$ so that the time is
of order $\Delta t_c\equiv t-\ti\simeq H^{-1}_{\rm out}(\ti) (c^{-1}\RW/
H^{-1}_{\rm out}(\ti))^2$.
If the void is much larger than the
Hubble radius outside the void ($c^{-1}\RW/$ $H^{-1}_{\rm out}(\ti)\gg 1$),
then the $1^{\rm st}$-crossing time is very large:
$\Delta t_c/H^{-1}_{\rm out}(\ti)\gg 1$, independent of the ``emptiness'' of
the void.
Other authors suggest
that the void would fill in with radiation.$^{\ref{LiddWan}}$
If spacetime at the void wall continues to expand as $a(t)$ when the
fluid diffuses into the void,
the comoving radius of the wall is roughly
$r\simeq r_0-c\int_\ti^t dt/a(t)$ or
\be
\label{eq:Rold}
ra(\ti)\simeq \RW-c\int_\ti^t dt' \sqrt{{\ti}/{t'}}
\ee
in a radiation-dominated universe.
The $1^{\rm st}$-crossing time (i.e. the earliest thermalization time)
then, is when $r\simeq 0$, or when
$\Delta t_c\simeq .5H^{-1}_{\rm out}(\ti)(c^{-1}\RW/\Hm)^2$,
which is roughly the same as the time for the horizon to ``engulf''
a conformally expanding void.\footnote{We thank Michael Turner for this
explanation.}
Because $R=ra(t)$, the radius of the void would be given by
$R\simeq(\RW-c\Hm\sqrt{t/\ti}~)\sqrt{t/\ti}$, which for nearly all of
the time is $R\simeq \RW\sqrt{t/\ti}$; the
void conformally stretches with spacetime.
At time $t\simeq \Hm (c^{-1}\RW/\Hm)^2$,
the void radius decreases quite rapidly to zero.
Thus, although the qualitative void evolution
is quite different in these two pictures, the quantitative $1^{\rm
st}$-crossing
times are not because in both the void comoves with spacetime.
The important point is
that the earliest possible thermalization time in both pictures
(i.e.~the $1^{\rm st}$-crossing time) is thought to be
\be
\label{eq:conf}
\Delta t_c\simeq H^{-1}_{\rm out}(\ti)(c^{-1}\RW/\Hm)^2.
\ee
We will show in this section that the actual
$1^{\rm st}$-crossing time is remarkably shorter than Eqn({\ref{eq:conf}}).
In doing so
we will show that Eqn({\ref{eq:Rold}}) is fundamentally flawed.

Consider two radially propagating photons A and B.  Photon A starts at
the inner edge of the void wall and propagates inward, and photon B begins at
the outer edge of the void wall and moves outward.
We would like to calculate the distance each travels in time $\Delta
t = t-\ti$, where $\ti$ is the initial time.
Using Eqns~({\ref{eq:metric}})
and ({\ref{eq:Gamlam}}), the infinitesimal coordinate distance traveled
by a
photon in time $dt$ is
$dr=cdt \Phi\Gamma/R'$.
Define $\rho_{\rm in}(t)$ and $\rho_{\rm out}(t)$ to be the energy densities
inside and outside the {\it evolving} wall region of the void.
(Thus the subscripts ``in'' and ``out'' refer only to the undisturbed fluid).
Initially, $\rho'_{\rm in}(\ti)=
\rho'_{\rm out}(\ti)=0$, and $\Gamma(\ti,R)=1$.  We also consider
non-viscous, relativistic fluids, so that $p=\rho/3$ and $Q(t,r)=0$.

We will first calculate the distances photons A and B travel in special
relativistic voids ($G_N=0$).  From Eqn.~({\ref{eq:dU}}), the fluid
acceleration is zero inside and outside the void: $\dU_{\rm in}=\dU_{\rm
out}=0$.
Therefore, $R(t,r)=R(\ti,r)$ inside and outside the void,
so that the infinitesimal distance traveled by photons A and B in time $dt$ is
$dR=c\Phi \Gamma dt$.  From Eqn.~({\ref{eq:drho}}), we see
that $\rho_{\rm out}$ and $\rho_{\rm in}$ are constant in time.
Since $\Phi_{\rm out}=1$, we find that
$\Phi_{\rm in}(t,r)=\Phi_{\rm in}(t_i,r)=(\rho_{\rm out}(\ti)/\rho_{\rm in}
(\ti))^{1/4}$ from Eqn.~({\ref{eq:Phirel}}).
In addition, since
$\dG\propto p'$ (Eqn.~({\ref{eq:dGexa}})), $\Gamma_{\rm in}(t)=
\Gamma_{\rm out}(t)=1$.
Therefore in time $\Delta t\equiv t-\ti$,
photon B travels outward the distance $\Delta R_{\rm B}(t)
\equiv R_{\rm B}(t)-\RW$ given by
\be
\label{eq:travel1}
\Delta R_{\rm B}=c\Phi_{\rm out}\Gamma_{\rm out}\Delta t=c\Delta t,
\ee
while photon A travels inward the distance $\Delta R_{\rm A}(t)
\equiv \RW-R_{\rm A}(t)$ given by
\be
\label{eq:travel}
\Delta R_{\rm A}(t)=c\Phi_{\rm in}\Gamma_{\rm in}\Delta t=
c\left[{\rho_{\rm out}(t_i)}/{\rho_{\rm in}(t_i)}
\right]^{1/4} \Delta t=c~{T_{\rm out}(\ti)}/{T_{\rm in}(\ti)}~\Delta t >
c\Delta t.
\ee
Thus the emptier the
void, the farther photon A moves
relative to photon B!
(It is important to note that this is a strictly relativistic effect; for
nonrelativistic voids (II-C),
$\Phi(\t,r)\simeq 1$ inside and outside this void so that
the distance traveled by photons A and B are
approximately the same:
$\Delta R_{\rm A}\simeq c\Delta t=\Delta R_{\rm B}$).
Define $\Delta t_c$ to be the $1^{\rm st}$-crossing time (the time taken for
photon A to reach the origin).
Then from Eqn.~({\ref{eq:travel}}) with $\Delta R_{\rm A}=\RW$, the
$1^{\rm st}$-crossing time is
\be
\label{eq:dtSR}
\Delta t_c=c^{-1}\RW\left[~{\rho_{\rm in}(\ti)}/{\rho_{\rm
out}(\ti)}~\right]^{1/4}.
\ee
Since $\rho_{\rm out}(\ti)\geq \rho_{\rm in}(\ti)$,
$\Delta t_c$ ranges from $\RW/c$ to zero.
Therefore in the limit that $\rho_{\rm in}(\ti)/\rho_{\rm out}(\ti)\rightarrow
0$,
a photon will reach the origin in {\it zero} cosmic time.
However, calling
a region a ``fluid'' if it is empty ($\rho_{\rm in}=0$) is incorrect.
Suffice it to say that
the $1^{\rm st}$-crossing time can be arbitrarily small.

We note that if Eulerian, synchronous coordinates were used with the metric
$ds^2=-c^2dT^2+dR^2+R^2d\Omega^2$,
the distance traveled by photon A or B in time $\Delta T$ would be the
same: $\Delta R=c\Delta T$.
Thus the time $\Delta T_c$ for photon A to reach the origin is
$\Delta T_c=c^{-1}\RW$.
(We can also see this using Lagrangian coordinates, since
the infinitesimal proper time measured by a comoving observer inside
the void is $d\tau=\Phi_{\rm in} dt=cdT$, and therefore
the time as measured by this observer for photon A to reach the origin is
$\Delta \tau_c=c\Phi_{\rm in}\Delta t_c=\RW$).
Thus, the quick $1^{\rm st}$-crossing time is
due to the choice of comoving, synchronous coordinates, a choice
we do not have for
superhorizon-sized general relativistic voids ($G_N\neq 0$)
embedded in a FRW universe.\footnote{
If the spatially flat FRW metric is transformed to the Eulerian gauge
with synchronous coordinates,
a coordinate singularity at the Hubble radius (see Appendix A).
Thus this gauge and coordinate choice cannot be used to describe the
evolution of superhorizon-sized general
relativistic voids.}

We now calculate the distance traveled by photons A and B in the same
cosmic time for a general relativistic void.
Again, the infinitesimal coordinate distance a photon travels in
time $dt$ is $dr=c\Phi\Gamma dt/R'$.
Because the pressure outside the void redshifts
due to Hubble expansion, $\Phi_{\rm in}(t)$ decreases in time from
Eqn({\ref{eq:Phirel}}) (since $\Phi_{\rm out}(t)=1$), and
$\Delta t_c$ consequently increases.
We first calculate the distance photon B travels.
Outside the void, $\Gamma_{\rm out}(t)=1$, and
$R=ra(t)$ so that
$R'=a(\ti)\sqrt{t/\ti}$. $^{\ref{Comp}}$
The comoving distance
photon B has traveled at time
$t$ is $\Delta r={c\ti}/{a(\ti)}(\sqrt{t/\ti}-1)$, so that
the distance photon B travels is
$\Delta R_B\equiv R_B-\RW=(\sqrt{t/\ti}-1)[\RW+1/2~c{H_{\rm out}^{-1}(\ti)}
\sqrt{t/\ti}]$.

We now calculate the location of photon A.
Assume that the energy density
inside the void changes negligibly:
$\rho_{\rm in}(t)=constant$.
(We will address
this approximation in a moment).
Then $\Phi_{\rm in}(t)=(\rho_{\rm out}(t)/\rho_{\rm in}(\ti))^{1/4}$.
Because the energy density outside the void is redshifted as
$\rho\propto 1/t^2$(see VI),$^{\ref{Comp}}$ $\Phi_{\rm in}(t)
\simeq\Phi_{\rm in}(\ti)\sqrt{\ti/t}$.
In addition, since
$\dG\propto p'$ (from Eqn.~({\ref{eq:dGexa}})), $\Gamma_{\rm in}(t)=
\Gamma_{\rm out}(t)=1$.
Integrating, we
find that the cosmic time taken for photon A to travel the distance
$\Delta R_A\equiv \RW-R_A$ is
\ba
\label{eq:DeltGR}
\Delta t\equiv t-\ti=c^{-1}\Delta R_A\left(\frac{\rho_{\rm in}(\ti)}
{\rho_{\rm out}(\ti)}\right)^{1/4}
\left[1+\frac{c^{-1}\Delta R_A}{2H_{\rm out}^{-1}(\ti)}
\left(\frac{\rho_{\rm in}(\ti)}{\rho_{\rm out}(\ti)}\right)^{1/4}\right],
\ea
and the location of photon A as a function of time is
\ba
\label{eq:radwint}
{R_A}&=&\RW-c\Phi_{\rm in}(\ti)\int_\ti^t dt'\sqrt{{\ti}/{t'}}\\
\label{eq:radwall}
&=&\RW-c\Phi_{\rm in}(\ti)\Hm\left[\sqrt{t/\ti}-1 \right].
\ea
As in the special relativistic case, $\Delta R_A(t)>\Delta R_B(t)$, so
that photon A travels farther than photon B in the same amount of
cosmic time.
The $1^{\rm st}$-crossing time relative to the initial outside Hubble
time is then ($\Delta R_A=\RW$)
\ba
\label{eq:DeltHGR}
\frac{\Delta t_c}{H^{-1}_{\rm out}(\ti)}=\frac{c^{-1}\RW}{H^{-1}_{\rm out}
(\ti)}\left(\frac{\rho_{\rm in}(\ti)}{\rho_{\rm out}(\ti)}\right)^{1/4}
\left[1+\frac{c^{-1}\RW}{2H_{\rm out}^{-1}(\ti)}
\left(\frac{\rho_{\rm in}(\ti)}{\rho_{\rm out}(\ti)}\right)^{1/4}\right].
\ea
In addition, if the more stringent condition
$c^{-1}\RW/H^{-1}_{\rm out}(\ti)<\Phi_{\rm in}(\ti)$ holds, then
\be
\label{eq:string}
\frac{\Delta t_c}{H^{-1}_{\rm out}(\ti)}\la 1~~~~~~~~~~~~{\rm when}
{}~~~~~\frac{c^{-1}\RW}{H^{-1}_{\rm out}(\ti)}<
\left(\frac{\rho_{\rm out}(\ti)}{\rho_{\rm in}(\ti)}\right)^{1/4}
\ee
---the minimum thermalization time (i.e. the $1^{\rm st}$-crossing time)
is less than the initial Hubble time outside the void!
(Note that the general and special relativistic results
(Eqns({\ref{eq:DeltHGR}})
and ({\ref{eq:dtSR}})) are equal in this case, since the energy density outside
the void is constant during this time).

Before discussing further implications, we find the condition for which
Eqn({\ref{eq:DeltGR}}) is satisfied; it
was derived under the assumption that
$\rho_{\rm in}(t)\simeq \rho_{\rm in}(\ti)$.
Using Eqn({\ref{eq:drho}}), inside the void
$|\drho/\rho|=4(R^2\dR)'/(3R^2R')=4\dR/R$ (since $R=ra(t)$) so that
if $\rho_{\rm in}\simeq constant$, then $R\simeq constant$ (in time) inside
the void.
Because
$U^2=(\dR/\Phi)^2=2G_NM/R\simeq c^{-2}(8\pi G_N\rho_{\rm in}/3)R^2$ inside
the void, the fractional change in the radius $R$
over time scale $\Delta t$ is roughly
\be
\label{eq:fin}
f\equiv \frac{\Delta R}{R}=\frac{\dot{R}\Delta t}{R}=\sqrt{\frac{\rho_{\rm in}
(t)}{\rho_{\rm out}(t)}}H_{\rm out}(t)\Phi_{\rm in}(t)\Delta t~
\simeq~\Phi^{-1}_{\rm in}(\ti)\sqrt{\frac{\ti}{t}}~
\frac{\Delta t}{H^{-1}_{\rm out}(\ti)},
\ee
where we have used the fact that $H^{-1}_{\rm out}(\ti)=2\ti$,
$\sqrt{\rho_{\rm in}(t)/\rho_{\rm out}(t)}H_{\rm out}(t)
\simeq\Phi^{-2}_{\rm in}(\ti) H_{\rm out}(\ti)$
and $\Phi_{\rm in}(\ti)=(\rho_{\rm out}(\ti)/\rho_{\rm in}(\ti))^{1/4}$.
We require $f<1$.
Writing $t=\Delta t +\ti$, Eqn({\ref{eq:fin}}) becomes
$\Delta t=f^2\Phi^2_{\rm in}(\ti){H^{-1}_{\rm out}(\ti)}[1+
\sqrt{1+1/(f\Phi_{\rm in}(\ti))^2}]$.
Since $\Phi_{\rm in}(\ti)> 1$, we have $f\Phi_{\rm in}(\ti)\ga 1$.
The condition for which the density inside of the void remains
approximately constant during time $\Delta t$ then, is
${\Delta t}/{H^{-1}_{\rm out}(\ti)}\leq 2f^2\Phi^2_{\rm in}(\ti)
\simeq \Phi_{\rm in}^2(\ti)$.
We now combine this with Eqn({\ref{eq:DeltGR}}) to find the maximum
allowed void size
$\Delta R_A/H^{-1}_{\rm out}(\ti)$ given $\Phi_{\rm in}(\ti)$.
We find $c^{-1}\Delta R_A/H^{-1}_{\rm out}(\ti)\leq .5\Phi_{\rm in}(\ti)[-1+
\sqrt{1+4(f\Phi_{\rm in}(\ti))^2}]\simeq f\Phi^2_{\rm in}(\ti)$.
Eqn~({\ref{eq:DeltGR}}) is then valid when
${c^{-1}\Delta R_A}/{H^{-1}_{\rm out}(\ti)}\leq f\Phi^2_{\rm in}(\ti)$,
and Eqn.~({\ref{eq:DeltHGR}}) is valid when
\be
\label{eq:rwvalid}
{c^{-1}\RW}~/~{H^{-1}_{\rm out}(\ti)}~\la~ \sqrt{{\rho_{\rm out}(\ti)}~/~
{\rho_{\rm in}(\ti)}}.
\ee
Note that Eqn({\ref{eq:string}})
is automatically satisfied.

The quick $1^{\rm st}$-crossing time might seem completely counterintuitive.
How can a photon travel a distance {\it much} larger than the Hubble radius in
less than a Hubble time?
The answer lies in describing how one measures
the size of an object which is {\it not} a small perturbation in spacetime.
If size is measured circumferentially,
then the void is enormous because its circumferential size
is $2\pi\RW\gg \Hm$.  (If spacetime were static, then it would take
a photon time $\Delta t\simeq 2\pi\RW/c$ to encircle the void).
However, if size is measured radially (the time taken for a
photon to cross the object if spacetime were static), then the void is measured
to be very small.
If fact, an interesting comparison can be made to measuring the size of a black
hole, an {\it overdense} region.
If one measures its circumferential size, it is small (or at least finite),
but its radial size is infinite.

The $1^{\rm st}$-crossing time for a superhorizon-sized void
was previous calculated incorrectly
because $t$ was assumed to be the proper time
outside {\it and} inside the void; in our
notation, it was implicitly assumed that $\Phi(t,r)=1$.
Comparing Eqns({\ref{eq:Rold}}) and ({\ref{eq:radwint}}),
we indeed see that the factor $\Phi_{\rm in}(\ti)>1$
is missing from Eqn({\ref{eq:Rold}}).  Note in addition that the factor of
$\sqrt{\ti/t}$ in Eqn({\ref{eq:radwint}})
does not come from spacetime expanding at the wall, but rather from the
density outside the void redshifting, causing
$\Phi_{\rm in}(t)$ to decrease.  In fact, spacetime is roughly non-expanding
at the inner wall edge.
If $c^{-1}\RW/\Hm > \Phi_{\rm in}(\ti) \gg 1$,
the actual position of photon A is approximately constant in time
until the last moment: $R_A\simeq \RW$ until $t\simeq \ti+\Delta t_c$
(see Eqn({\ref{eq:radwall}})).
Thus, spacetime at the inner edge of the void wall is neither expanding nor
contracting.

We note the
interesting fact that if a photon inside the void is in thermal equilibrium
with average frequency $\nu$, its
frequency is $T_{\rm in}$.  If it
moves outside the void, its frequency is blue-shifted
to $\nu'=\nu\Phi_{\rm in}=\nu (T_{\rm out}/T_{\rm in})=T_{\rm out}$, the
average
frequency of thermal photons outside the void.
Thus, this photon is automatically in thermal equilibrium outside
the void, so that an outside observer
could not detect the void's initial presence unless non-thermal
photons came out.\footnote{Because entropy is
created at the shock (see VII), the photon with $\nu T_{\rm in}$ would actuall
have a slightly
higher frequency than thermal outside the void.}
This is essentially because the fluid is relativistic, or that
$\rho$, $p$, $n$ and $\ep$ depend only on the
temperature (II-C).
We can understand why the photon is blue-shifted by way of comparison
to an (overdense) black hole.  Suppose
an observer falls into a black hole ticking off photons at a fixed frequency.
As this observer crosses the event horizon, the frequency of the
photon emitted last gets redshifted to infinity as observed by a
stationary observer
outside the black hole.$^{\ref{LLfluids}}$  The opposite effect happens for
a photon emitted from an underdense region.  Because a photon leaving
a void enters a region with a much larger gravitational potential,
the frequency instead gets blue-shifted.

In conclusion, the $1^{\rm st}$-crossing time for a superhorizon-sized
relativistic void embedded in a FRW expanding universe
is given by Eqn~({\ref{eq:DeltHGR}}) (if Eqn({\ref{eq:rwvalid}}) is satisfied),
and depends sensitively on the quantity
$\RW/H^{-1}_{\rm out}(\ti)$ $(\rho_{\rm in}(\ti)/\rho_{\rm out}(\ti))^{1/4}$.
We emphasize the important point that if
$c^{-1}\RW/H^{-1}_{\rm out}(\ti)<(\rho_{\rm out}(\ti)/\rho_{\rm in}(\ti))^{1/4}
=T_{\rm out}(\ti)/T_{\rm in}(\ti)$,
then the $1^{\rm st}$-crossing time is less than the outside
Hubble time: $\Delta t_c/H^{-1}_{\rm out}(\ti)<1$.

\centerline{\bf{V. Pressureless Non-Viscous Voids}}
\setcounter{section}{5}
\setcounter{equation}{0}

For the special case when the pressure and viscosity are zero,
the equations of motion can be solved analytically.  These
give the Tolman-Bondi dust solutions,$^{{\ref{TolBon}},{\ref{LLfluids}}}$
which we will review here briefly.
It is important to study the pressureless case not only as a test problem, but
also to see how removing fluid forces affects the
evolution of a void.  (Because $p=Q=0$, the particles move only under
gravitational
forces: $\Phi^{-1}\dU=-{G_NM}/{R^2}$ (see Eqn~({\ref{eq:dU}}))).

Since $\Phi'=0$ (Eqn~({\ref{eq:defPhi}})), we set $\Phi(t,r)\equiv 1$.
The mass contained within $r$ remains constant:
$M(r)=\int_0^r 4\pi n R^2 dR/\Gamma$, since
$\dM=0$ (see Eqn~({\ref{eq:dM}})).  And from Eqn({\ref{eq:dGexa}}),
$\Gamma(t,r)=\Gamma(\ti,r)$.
For $R'\neq 0$, $\rho=n$ can be found from Eqn({\ref{eq:defn}}):
$\rho(t,r)=\rho(\ti,r)~{R(\ti,r)'R(\ti,r)^2}/[{R(t,r)'R(t,r)^2}]$.
The generalized pressureless Friedmann equation,
Eq.~(${\ref{eq:defGamma}}$), now becomes
\be
\label{eq:tolbon}
\dR^2=c^2\left[\Gamma(r)^2-1\right]+2G_N M(r)/R.
\ee
The quantity $U^2/2-G_NM/R=c^2(\Gamma^2-1)/2$ is conserved during evolution
and can be interpreted as the generalized total energy.  ~Eqn
({\ref{eq:tolbon}}) can be integrated for a shell of radius
$\r$:
\ba
\label{eq:TolBon}
R&=&\frac{c^{-2}G_NM}{\Gamma^2-1}\left(\cosh\eta-1\right),~~~~
t=\tau_0(r)+\frac{c^{-3}G_NM}{(\Gamma^2-1)^{3/2}}\left(\sinh\eta-\eta\right),~~~
{\rm {for}}~\Gamma(r)^2>1  \nonumber\\
R&=&\frac{c^{-2}G_NM}{1-\Gamma^2}\left(1-\cos\eta\right),~~~~
t=\tau_0(r)+\frac{c^{-3}G_NM}{(1-\Gamma^2)^{3/2}}\left(\eta-\sin\eta\right),~~~
{\rm {for}}~\Gamma(r)^2<1  \nonumber\\
R&=&\left({9G_NM}/{2}\right)^{1/3}\left(t-\tau_0(r)\right)^{2/3},~~~~{\rm
{for}}~
\Gamma(r)=1.
\ea
These are the Tolman-Bondi solutions.

In these models, shell-crossing can occur.  This happens
when two adjacent shells (labeled by $r$ and $r+dr$) occupy the same
position so that
$R'=0$ and $\rho\rightarrow\infty$.
This may lead to a non-unique continuation of the solution,
$^{\ref{Newm}}$ and thus
computations have to be stopped.
This problem is believed
to occur
because the pressure has been artificially set to zero.
It is generally thought that adding pressure
would prevent this situation from occurring.\footnote{The author thanks T.
Piran
for this information.}
As will be seen, when $\Gamma(r)> 1$ in the void wall region,
shell-crossing does occur. The addition of enough artificial viscosity
can prevent this from happening, however.
Even though the viscosity given by Eqn~({\ref{eq:Visc}}) was designed to
stabilize numerical shocks,
it has been found to prevent shell-crossing.

For the numerical simulations in this section, we set $G_N=1$, $c=1$, $\ti=1$,
$\rho(\ti,R)=0$, $\ep(\ti,R_B)=0$, $C=.3$, $\f=.005$ and $\gamma=5/3$.
Thus, the initial energy
density and Hubble radius outside the void are
$4\pi\rho_{\rm out}(\ti)=2/3$ and $H^{-1}_{\rm out}(\ti)=3/2$, respectively
(see III-A).

We first examine the situation in which
each mass shell's velocity initially compensates
for the gravitational attraction inwards: $\Gamma(\ti,R)=1$.
Then $U(\ti,R)=\sqrt{2G_N M/R}$.
Figure 1 shows the energy density versus $R~R(\ti,R_B)/R(t,R_B)$
($\equiv R~R_{j_B}(\ti)/R_{j_B}(t)$)
for a superhorizon-sized
compensated void with $c^{-1}\RW/H^{-1}_{\rm out}(\ti)=333$,
$\rw=15$, $\alpha=.001$, $\DR=2.5$
and $k^2=0$.
We show the analytic and numerical results at
times $\t=1,~10,~100,~{\rm and~}300$ where $R_{j_B}(t)=1002,~4651,~2.16
\times 10^4,~{\rm and}~4.49\times 10^4$ respectively.
The triangles and squares are the numerical and Tolman-Bondi solutions,
respectively,
although they are difficult to distinguish because the numerical results
agree so well with the analytic results.
By $t\sim 300$, the density everywhere
is approximately constant; there is hardly a trace of the void's initial
presence.
Identical results are obtained for subhorizon-sized voids.
In addition, an initially uncompensated
void evolves similarly.

In Table 1, we show the results of a convergence test for $q=\rho$
applied to the same initial conditions as in Figure 1, but for variable $\DR$
(see III-C).
(We only do this test for $\rho$, because the accumulated error in $M$ and $R$
are much smaller).
We set up analytic conditions initially and integrate until $t=1.5$.
Because the inner grid point or two ends up being the numerical culprit
for non-second order convergence, we also calculate
$L_{1a}\equiv\frac{1}{N-2}\sum_{i=3}^N e_i$ and $L_{1b}\equiv\frac{1}{N-10}
\sum_{i=10}^N e_i$,
where $e_i$ is the relative error.
(This is because the code consistently underestimated
$\rho$ at the innermost few grid points).
To the right of each global error estimate, the convergence
rate is shown ($L_i\propto \DR^s$).
We see that the convergence rate for $L_1$ is
less than second-order, whereas
that for $L_1$ is nearly second order.
\begin{table}[t]
\begin{center}
\caption{Convergence test for Tolman-Bondi model}
\begin{tabular}{|c|c|c|c|c|c|c|}
\hline
\hline
\multicolumn{1}{|c|}{$\DR$}&\multicolumn{1}{|c|}{$L_1$}&\multicolumn{1}{|c|}{s}
&\multicolumn{1}{|c|}{$L_{1a}$}&\multicolumn{1}{|c|}{s}&\multicolumn{1}{|c|}
{$L_{1b}$}&\multicolumn{1}{|c|}{s}\\
\hline
\hline
$8$&$ 1.2\times 10^{-3} $&$ ... $&$ 8.33\times 10^{-4} $&$ ... $&$ 8.55\times
10^{-4} $&$ ... $\\
\hline
$4$&$ 4.13\times 10^{-4} $&$ 1.48 $&$ 2.26\times 10^{-4} $&$ 1.89 $&$
2.16\times
10^{-4} $&$ 2.0 $\\
\hline
$2$&$ 1.55\times 10^{-4} $&$ 1.32 $&$ 6.09\times 10^{-5} $&$ 1.82 $&$
5.35\times
10^{-5} $&$ 1.97 $\\
\hline
$1$&$ 6.56\times 10^{-5} $&$ 1.24 $&$ 1.81\times 10^{-5} $&$ 1.75 $&$
1.41\times
10^{-5} $&$ 1.92 $\\
\hline
\hline
\end{tabular}
\end{center}
\end{table}

We have just seen that if $U=U_{\rm GRAV}$ ($\Gamma(\ti,R)=1$), the void
disappears.
What happens when $U>U_{\rm GRAV}$ ($\Gamma(\ti,R)>1$) in the wall region?
Recall that this  corresponds to a net outward
peculiar velocity (II-D).
Figure 2 shows the result for a compensated void with
$c^{-1}\RW/H^{-1}_{\rm out}(\ti)=333$ ,
$\rw=15$, $\alpha=.001$, $\DR=2.5$,
and $k^2=0$.
We choose the initial velocity to be given by
Eqn({\ref{eq:velbig}}).  Therefore, $\Gamma(\ti,R)=1$ everywhere except in
the wall region, where $\Gamma(\ti,R)>1$.
Again, the triangles and squares represent the numerical
and analytic solutions, respectively.  The initial time ($\ti$) and
$t=1.02,~1.048$ are shown.
Again, the difference between the numerical and analytical results
are small except at $t=1.048$, where shell crossing occurs in both
solutions, a good check on the code.
The comoving radius of the shell with the highest density,
$r_{\rm shell}(t)$, remains
approximately constant in time.
Because each shell in the wall has constant total energy
$c^2(\Gamma^2-1)/2$, the wall expands outward.
(Since $U>0$, $R$ increases.  But since the total energy is constant
and $M(r)$ is constant, $U$ must increase).
Identical results are obtained for subhorizon voids.

Figures 3a and 3b show the density as a function of $R~R(\ti,R_B)/R(t,R_B)$ for
a subhorizon-sized, shallow, uncompensated and compensated void, respectively,
at $t=1,~50,~100,~400~{\rm and~}1000$.  Note that shell-crossing would
have occurred at $t=116$ and $2.1$ for the uncompensated and compensated
voids, respectively.
Here, $c^{-1}\RW/H^{-1}_{\rm out}(\ti)=.0067$, $\rw=.001$, $\DR=.0001$,
$\alpha=.5$, and
the velocity $U(\ti,R)$ is given by Eqn~({\ref{eq:velbig}}).  In addition,
for Figure 3a,
$k^2=4$ and $R_{j_B}(t)=.035,~.49,~.79,~2.13~{\rm and}~4.26$, while
for Figure 3b, $k^2=8$ and
$R_{j_B}(t)=.035,~.48,~.76,~1.9~{\rm and}~3.5$.
As can be seen, the initial perturbation
grows with time and eventually forms a thin, dense shell.  Again,
the comoving coordinate for the shell with the highest density
is approximately constant in time after the shell has formed.
As the shell travels outward, it pushes mass in front of it, producing
a shock.  This
situation is similar to that of a fast car
colliding with slower ones; although the
faster car is not allowed to move through the slower cars, momentum is still
transferred to them.
Note that the initially compensated perturbation forms a thick shell
more quickly than the uncompensated perturbation, although
it then proceeds to grow more slowly.

As long as the voids formed remain subhorizon-sized, they will eventually
grow according to a known similarity solution.$^{\ref{Bert}}$
An initially compensated (uncompensated) perturbation
in an expanding FRW $p=0$ universe will eventually form a dense, thin shell
that expands outward as $R_{\rm shell}\equiv R(t,r_{\rm shell})
\propto t^{4/5}$ ($t^{8/9}$).
In Figure 3c, we show the position of the void wall versus time for
the results of Figures 3a and 3b.
The triangles and squares connected by lines are the
numerical solutions for the compensated and uncompensated
cases, respectively, and the dashed and dotted lines
are the self-similar solutions for the compensated and uncompensated cases,
respectively.
As can be seen, the initially compensated perturbation
approaches the similarity solution more quickly than the initially
uncompensated
perturbation, but for $t\ga 800$, both solutions are self-similar.

\centerline{\bf{VI. FRW Homogeneous Cosmologies}}
\setcounter{section}{6}
\setcounter{equation}{0}

We now test our code
against the exact FRW homogeneous and isotropic solution
for relativistic fluids with $p=\rho/3$.
As discussed in II-A,B and III-A, the FRW solution is
$R(t,r)=ra(t)=R(\ti,r)(t/\ti)^\xi$ where $\xi=1/2$.
In addition, $4\pi G_N\rho(\t)=3c^2/(8t^2)$ and $U=R/(2t)$.
For the numerical results shown in this section, we set $G_N=1$, $c=1$,
$\gamma=4/3$, $C=.3$, $\ti=1$,
$\ep(\ti,\Rmx)=10^6$
and $c^{-1}R_B(\ti)/H^{-1}(\ti)=250$ so that $4\pi\rho(\ti,\Rmx)=3/8$.

Figure 4 shows the relative error in $\rho$ (III-C)
for a simulation with
$\DR=.5$ and $k^2=0$.
The analytical solution was set up initially, and the code was
run until $t=1.1$.
The solid, dotted and dashed lines are for $\f=.01$, $\f=.005$
and $\f=.0025$.  The relative error at the
outer boundary is seen to be very sensitive to $\f$;
it is $.05\%$, $.15\%$
and $2\%$ for $\f=.0025$, $.005$ and $.01$ respectively.
Note also the underprediction of $\rho$ at the innermost few grid points.

Table 2 shows the accumulated error at time $t=1.1$
for simulations with $k^2=4$. \footnote{When
evolving voids, viscosity is absolutely
necessary.  It is therefore important to see how it affects the solution
where the fluid is approximately homogeneous and isotropic.
It turns out that it
is virtually unaffected by $Q\neq 0$, as it should be.}
Again, $L_{1b}\equiv\frac{1}{N-10}\sum_{i=10}^N e_i$,
where $e_i$ is the relative error
for $q=\rho$.
In the first $7$ columns, we show the results for
$N=25,~50,~100,~200,~400,~800$
and $1600$ (i.e. $\DR=N/250$)
and for $\f=.01,~.005,~{\rm and~}.0025$.
For a given value of $\f$, as $N$ increases, ~$L_{1}$ and
$L_{1b}$ approach the same constant value even though $L_1$ starts
out much larger.  After this constant value has been reached,
$L_1$ and $L_{1b}$ remain unchanged when $\DR$ is further decreased,
even though the relative error near the origin improves.
This is because beyond a certain point, all the
accumulated error comes from the outer boundary, which is immune to changes
in $\rw$ (see III-B).

Knowing that a given value of $\f$ limits
convergence of the code, we can
calculate the convergence as a function of
the asymptotic value of $L_1$.  We assume
that $L_1(\DR\rightarrow 0)\propto \f^\s$.
Column $9$ gives the estimated value for $L_1$($\DR\rightarrow 0$),
and column 10 estimates the value of $\s$.  We see that
$\s$ is nearly $2$, which means that convergence in $\f$
is nearly second order given a small enough value for $\DR$.

\begin{table}[t]
\begin{center}
\caption{Ultrarelativistic Homogeneous convergence test}
\begin{tabular}{|c|l|l|l|l|l|l|l|l|l|l|}
\hline
\hline
\multicolumn{1}{|c|}{ }&\multicolumn{7}{|c|}{Number of grid points (N)}
&\multicolumn{1}{|c|}{ }&\multicolumn{1}{|c|}{$L_1$}&\multicolumn{1}{|c|}{ }\\
\cline{2-8}
 &$\;\;\; 25$&$\;\;\; 50$&$\;\; 100$&$\;\; 200$&$\;\; 400$&$\;\; 800$&$ 1600$&
$\;\;\;\f$&{\footnotesize{$\Delta R\rightarrow 0$}}&$\;\; \s$\\
\hline
\hline
$L_1$&$ .016$&$ .010$&$ .0073$&$ .0058$&$ .0050$&$ .0046$& &$.01$&$.004$&...\\
\cline{1-8}
$L_{1b}$&$.0071$&$.0056$&$.0049$&$.0045$&$.0044$&$.0043$&$ $& & &\\
\hline
$L_1$&$.012$&$.0067$&$.0038$&$.0023$&$.0016$&$.0012$&$ $&$.005$&
$.00085$&$2.00$\\
\cline{1-8}
$L_{1b}$&$.0015$&$.0012$&$.0010$&$.00091$&$.00088$&$.00087$& & & \\
\hline
$L_1$&$.011$&$.0059$&$.0031$&$.0017$&$.00096$&$.00060$&$.00042$&$.00025$&
$.00025$&$1.77$\\
\cline{1-8}
$L_{1b}$&$.00057$&$.00041$&$.00032$&$.00027$&$.00026$&$.00025$&$.00025$& & &\\
\hline
\hline
\end{tabular}
\end{center}
\end{table}
\normalsize

\centerline{\bf{VII. Numerical Evolution of Voids}}
\setcounter{section}{7}
\setcounter{equation}{0}

\centerline{\bf{A. Nonrelativistic Fluids}}

In this subsection we examine the evolution of voids composed of
nonrelativistic
particles in zero gravity ($G_N=0$).  If $T$ is the fluid temperature
and $\mu$ is a fluid particle's mass,
$T/\mu\ll 1$.
Since $H^{-1}\propto {G_N}^{-1}$,
these voids are subhorizon-sized.
For the simulations in this subsection,
we set $G_N=0$, $c=10^{10}$, $\ti=1$, $C=.3$, $\gamma=5/3$, $C=.3$,
$4\pi\rho(\ti,R_B)=2/3$ and $\ep(\ti,R_B)=1$.

We start with the shock tube problem, a standard test of
1-D slab codes.$^{{\ref{Sod}}}$
In addition, it provides insight into the dynamics of
collapsing voids.
In a shock tube, the fluid is initially at rest and is separated into two
regions with different pressures
and densities.
The pressure discontinuity produces a shock wave which propagates into the
low pressure region and a rarefaction wave which propagates into the
high pressure region.
An analytic similarity solution exits for a perfect fluid with slab geometry.
It does not exist for the spherically
symmetric geometry however.$^{\ref{Sedov}}$
Far from the origin however, the spherically symmetric solution approaches
the slab solution for small times and distances.$^{\ref{MayWhite}}$
We will thus set up a spherically symmetric shock tube by evolving
an uncompensated void far from the origin, and compare the results to the
exact slab similarity solution (briefly reviewed in Appendix D).

In Figure 5 we show the results for the shock tube problem
with $U(\ti,R)=0$ (or $\Gamma(\ti,R)=1$),
$\RW=20$, $\rw=.01$, $\DR=.01$, $\alpha=.01$
and $k^2=5$.
(We do not take the pressure gradient to be discontinuous initially,
because the solution
in the original wall area is not as accurate then.\footnote{This is primarily
because we calculate $n$ via
$\dn\propto (R^2dR)^{-1}$ rather than $\dn\propto (r^2dr)^{-1}$})
We plot the pressure, number density, velocity
and specific energy as a function of the position $R$
at $\ti$ and at $t=1.15$.
The triangles connected by lines is the numerical solution, and the
dashed lines is the slab similarity solution.
At $t=1.15$, a strong shock wave (located at $R\simeq 19.68$) moves inward
and a rarefaction wave (between $19.85\la R\la 20.14$) moves outward.
Note that the shock is spread out over $k^2\simeq 4-5$ grid points.
The numerical and analytical solutions are seen to
agree well in this limit, because the spherical
geometrical effects are small ($(R_{shock}(t)-\RW)/\RW\sim .4/20=.02$).
The distortion of the velocity distribution is due to the small geometrical
effect;
the shock gets slightly
stronger directly behind the shock due to the
smaller effective volume $4\pi R^2\Delta R$ those mass shells occupy
relative to shells further back.
An important point to emphasize is that although initially the fluid is
everywhere
stationary ($U(\ti,R)=0$), it acquires a net momentum to the left in the wall
region.

In Figure 6a we show the long-term results for an initial configuration
with $\RW=1$ but otherwise identical to Figure 5.
The pressure, number density, velocity and specific energy
are shown at the initial time $\ti=1$ with dashed lines, and long after
the collision at $t=2.5$ with triangles
and connecting lines.
Although it is not shown for clarity, the fluid configuration before the shock
rebounds at the
origin is similar to Figure 5:  a shock
heads toward the origin and a rarefaction wave moves away from the origin.
When the (spherical) shock crashes into the origin,
the volume effect proves harsh as the fluid collides with itself
in a vanishingly
small volume.  This causes the pressure at the origin to become
very large in order to repel the fluid (not pictured here).
Note that the fluid at the shock as well as far behind it
must be repelled, since it is all moving toward the origin.\footnote{It is this
reversal that requires a very robust difference scheme.  All
of the ``obvious'' difference schemes failed after the inbound shocks
reached the
origin.  Setting $f=r^2$ in Eqn~({\ref{eq:deff}}) and using
those difference schemes listed in Appendix B are the very necessary
requirements.}
When the dust settles, we find that a weak shock has rebounded back.  This
outward-moving shock can
be seen at $R($t=2.5$)\simeq 1.3$.  Due to the volume effect however, this
shock will become weaker as it moves outward further.
Also seen at $t=2.5$ is the original
(outgoing) rarefaction wave located between $1.3\la R\la 2.4$.
Note that at $t=2.5$, the fluid is (and will approximately remain)
at rest near the origin
because the velocity is zero and the pressure is constant.
However, a large distortion has been left behind in the fluid in the form of
low density and high kinetic energy.  This consists entirely
of the fluid originally in the void.
The low-$n$, high-$\ep$ values for the
first 5-6 grid points, however, is artificial.  This is a consequence
of using VonNeumann-type artificial viscosity called ``wall-heating'',
and is caused by the collision between 2 shocks.$^{\ref{WallHeat}}$

In Figure 6b we show the initial and long-term pressure, number density,
velocity and specific energy as a function of the radius
for a compensated, nonrelativistic
void.  The initial distribution is identical to Figure 6a except for
the energy density, and $\rw=.02$.  Again, the fluid configuration at
$\ti=1$ is shown as dashed lines and that at $t=2.5$ is shown as
triangles connected by lines.
An inbound shock is again formed from the initial pressure gradient at
the inner edge of the void wall.
Unlike the uncompensated case however, a weak outgoing shock wave is formed
instead
of a rarefaction wave.  At $t=2.5$, it is located at $R=3.2$.
Like the uncompensated void, the pressure at the origin becomes very large
after the shock
collides with itself there, and a weak shock rebounds.
At $t=2.5$, this rebounded shock is located at $R\simeq 2.3$, which is
farther out than that for the uncompensated void ($R\simeq 1.3$).
This is because the inbound shock produced from
the compensated case is much stronger than for the uncompensated case.
And again there is a large distortion left
near the origin containing all the fluid initially in the void,
although it is somewhat different spatially.

We now compare the collapse times for uncompensated and compensated voids of
varying wall thicknesses.
(The collapse time is defined to be the time taken for the shock to reach the
origin).  We set
$\RW=1$, $\alpha=.01$, $k^2=3$ and $\DR=.01$.
Table 3 shows the results for
$\rw=.02$,~.04$,~.08$ and $.12$, where we calculate the percent
change by comparing the collapse time with the uncompensated $\rw=.02$ collapse
time of $\Delta t=.38$.
As $\rw$ increases, the collapse time
increases.  In addition, the collapse time for compensated voids is
substantially smaller than for uncompensated voids.

\begin{table}[t]
\begin{center}
\caption{Nonrelativistic collapse times}
\begin{tabular}{|l|l|l|l|l|}
\hline
\hline
\multicolumn{1}{|c|}{Void type}&\multicolumn{1}{|c|}{$\rw=.02$}&
\multicolumn{1}{|c|}{$\rw=.04$}&\multicolumn{1}{|c|}{$\rw=.08$}&
\multicolumn{1}{|c|}{$\rw=.12$}\\
\cline{2-5}
\multicolumn{1}{|c|}{ }&\multicolumn{4}{|c|}{Collapse Time $\Delta t_c$ ~~$||
$~~ Percent Change: $(\Delta t_c-\Delta t_c(.02)_{\rm un})/
\Delta t_c(.02)_{\rm un}$}\\
\hline
\hline
uncompensated&$ .38~~||~~~~~...$&$ .41~~||~~~~~~8\%$&$ .45~~||~~~~~18\%$&
$ .50~~||~~~~~32\%$\\
\hline
compensated&$ .15~~||~-61\%$&$ .20~~||~-47\%$&$ .27~~||~-29\%$&$
.32~~||~-16\%$\\
\hline
\hline
\end{tabular}
\end{center}
\end{table}

In conclusion, nonrelativistic voids
with zero gravity collapse in the form of
a shock, the strength of which depends on the details of the void wall.
Some time after collapsing, the fluid is virtually at rest everywhere, with
$n$ and $\ep$ inhomogeneous near the origin.

\centerline{\bf{B. Special Relativistic Fluids}}

In this subsection we consider the evolution of special relativistic voids
with $T/\mu\gg 1$, where $T$ is the temperature and $\mu$ is the mass
of a fluid particle.
Thus, $\ep/c^2\gg 1$, and we set $G_N=0$, $c=1$, $\ti=1$, $C=.3$,
$\gamma=4/3$, $\ep(\ti,R_B)=10^3$, $U(\ti,R)=0$ (or $\Gamma(\ti,R)=1$) and
$4\pi\rho(\ti,R_B)=3/8$ in this subsection.

In Figure 7a, we show $M$, $\Gamma$, $4\pi\rho$ and $\Phi$ as a function
of $R$ for a relativistic shock tube problem.
Here $\RW=1$, $k^2=3$, $\rw=.02$, $\DR=.01$ and $\alpha=10^{-4}$.
As in the nonrelativistic
case, an inbound shock is formed.
(This is observed most easily in the plot of $4\pi\rho$ versus $R$).
However, an outgoing ``rarefaction wave'' is not observed, as it is in the
nonrelativistic case (see Figure 5).
If a wave were to propagate outward, it could go at most the
distance a photon would travel.
But from Eqn({\ref{eq:travel1}}), we know that a photon starting from
the outer wall edge only moves the distance $\Delta R_B=.04$ ($.065$)
in time $\Delta t=.04$ ($.065$).
Since this is of order
the grid point thickness, if an outbound wave were present, it would not be
observed at this time anyway.
On the other hand,
an inbound photon starting from the inner wall edge
would travel the distance $\Delta R_A=.4$ ($.65$) (from Eqn({\ref{eq:travel}}))
in time $\Delta t=.04$ ($.065$), since
$\rho_{\rm out}(\ti)/\rho_{\rm in}(\ti)\simeq 10^4$.
Reexamining Figure 7a, we now notice an
important result; the inbound shock's position is approximately equal to
photon A's location---the shock moves inward at roughly the speed of
light.  This is actually not so surprising, because
the speed of sound for a perfect fluid with $p=\rho/3$ is
$.57c$ (III-C).

Consider next voids compensated in energy density.
Because of relativistic-particle diffusion, we expect
the inbound shock to again travel at approximately the speed of light.
Since the value of $\Phi_{\rm in}(\ti)$
does not depend on the functional form of $\rho$
in the void wall (Eqn~({\ref{eq:Phirel}})),
the shock should move the same distance per time
as for the uncompensated void.
We ran
a compensated void simulation with the same initial conditions as from
Figure 7a
(except for $\rho$ in the void wall), and found
that at $t=1.04$ and $1.065$, the compensated
shock is ahead by only $\Delta R=.05$.  However, this can be accounted for by
the slightly different values of $\rho$ initially at the inner
edge of the wall.
Thus, special relativistic compensated and uncompensated voids collapse
at approximately the same speed, unlike nonrelativistic voids (see Table 3).

In Figure 7b, we show the numerical results for an uncompensated void  with
$\RW=1$, $k^2=3$, $\alpha=10^{-4}$ and $\DR=.01$, and
for $\rw=.02$, $.04$, $.06$ and $.1$.
It is clear from this figure that in all four cases
the shock reaches the origin
at $\Delta t_c\sim .08$; $\Delta t_c$ is approximately independent of $\rw$.
(Compare this with Table 3).
Using Eqn~({\ref{eq:travel}}),
we estimate the time for light to reach the origin at
$\Delta t_c\simeq \RW/\Phi\simeq .09$, in agreement with the
simulations.\footnote{Because $\Phi_{\rm in}$ actually increases
slightly due to the
dissipation of energy at the shock, the actual travel time for the shock
to reach the origin,
$\Delta t_c$, is decreased slightly.}
(During this time, photon B would only move
outward the distance $\Delta R_B= \Delta t_c\simeq .09$).
The value of $\rw$ however, does influence shock formation-time and strength.
As $\rw$ increases, the shock formation time increases and
the shock strength decreases.

We now examine the what happens to the void after the shock collides at
the origin.
In Figure 7c, we show the numerical results for an uncompensated void
with $\RW=1$, $\rw=.02$, $\alpha=10^{-4}$, $k^2=8$ and $\DR=.01$.
\footnote{After
colliding, a very weak shock rebounds.
The generalized functional form for the artificial viscosity, however,
was derived in the strong shock limit.
Thus, a
large value of $k^2$ was needed to maintain numerical stability after the
collision.
A new functional form for $Q$ will have to be used in future
simulations to stabilize the weak outgoing shock.$^{\ref{Vasha}}$}
We show $p,~n,~U~{\rm and~}\ep$ as a function of $R$
at the initial time $\ti$ and at times $1.04~{\rm and~}3.0$, where the second
and third times are before and
after collision at the origin, respectively.
At $t=3$, $p'\simeq 0$ and $U'\simeq 0$ near the origin; the fluid there is
roughly at rest.
Two weak outgoing waves are observed: the shock (at $R\sim 1$) and a
``rarefaction wave'' (between $1.1\la R\la 2.1$).
In addition, in contrast with nonrelativistic voids,
$n'\sim 0$ and $\ep'\sim 0$
near the origin.\footnote{For
the innermost 7-8 grid points, $\ep$ and $n$ are too large and too small,
respectively.
This is again due to ``wall-heating''.}
This result
is expected; since
$n\propto p^{3/4}\propto T^3$ and $\ep\propto p^{1/4}\propto T$
(from the discussion following
Eqns~({\ref{eq:isentrop}})) for a non-viscous fluid,
if $p'\simeq 0$, then it follows that $n'\simeq 0$ and $\ep'\simeq 0$.
This is a consequence of the fact that $\rho$, $p$, $n$ and $\ep$
depend only on the temperature for $\ep/c^2\gg 1$ and $Q=0$.
We therefore find that after the collapse, there is only a small trace of the
void's initial presence!

We conclude that a special relativistic void collapses in the form of a shock
which travels at approximately the speed of light into the void.  Thus, the
collapse
time is
of order the $1^{\rm st}$-crossing time.
Some time after the collapse, the fluid becomes approximately homogeneous and
isotropic everywhere.

\centerline{\bf{C. General Relativistic Fluids}}

In this subsection we study the evolution of general relativistic voids for
$T/\mu\gg 1$, where $T$ is the temperature and $\mu$ is the mass
of a fluid particle.
We set $G_N=1$, $c=1$, $\ti=1$, $C=.3$ and
$\gamma=4/3$.  Therefore, the outside Hubble radius and density
(that outside the void) are $cH^{-1}_{\rm out}(\ti)=2$ and
$4\pi G_N \rho(\ti,R_B)=3/8$.

In Figure 8, we show the pressure, number density, velocity and specific
energy for a general relativistic
void at the initial time $\ti$ and for $t=1.04$ and $t=1.065$.
The initial conditions
are identical to those of Figure 7a, except that $k^2=4$ and
$U(\ti,R)=U_{\rm GRAV}=\sqrt{2G_N M/R}$ (or $\Gamma(\ti,R)=1$).
Because the void is subhorizon-sized ($c^{-1}\RW/H^{-1}_{\rm out}(\ti)=1/2$),
its
evolution looks virtually the same as that for the special relativistic void
shown
in Figure 7a; at the void wall, the gravitational force is
$M/R^2+4\pi pR/c^2\simeq 8\pi\rho R/3\simeq 10^{-1}$, which
is much less than the fluid force $\Gamma^2p'/(4pR')\simeq (4\DR)^{-1}\simeq
12$
(see Eqn.~({\ref{eq:dUmss}})).

In Figure 9a, we show the pressure as a function of
$R~R(\ti,R_B)/R(t,R_B)$ ($\equiv R(t,r)$ $R_{j_B}(\ti)/R_{j_B}(t)$)
for a general relativistic superhorizon-sized void at $\ti$, $t=2.0$ and
$t=8.0$
with $\Gamma(\ti,R)=1$,
$\RW=50$, $\rw=1.$, $k^2=4$, $\alpha=10^{-4}$, $\ep(\ti,\rmx)=10^{3}$ and
$\DR=.5$.
(This void is $50/2=25$ times the outside Hubble radius).
In addition, $R_{j_B}(t)=100.2,~139.1,~{\rm and}~271.8$.
Even though $G_N \neq 0$ here, a strong inward shock still forms.  This
is because particles are diffusing into the void, having been accelerated
away from the high-pressure wall.
(Note that at the wall, the acceleration due
to gravity is $G_NM/R^2+4\pi pR/c^2\simeq 10$,
whereas that due to the fluid force is only $\Gamma^2p'/(4pR')\simeq .25$.
This small relative amount however, is enough to form the shock).
Note that because of expansion, the pressure outside (and
to a lessor extent inside) the void redshifts.  By
$t=8$, the pressure outside the void has redshifted from $p\simeq 10^{-1}$ to
$2\times 10^{-3}$, the expected amount since $p\propto 1/t^2$ so that
$p_{\rm out}(8)\simeq 10^{-1}/64\simeq 2\times 10^{-3}$.
At the same time, the pressure
on the inside has
only redshifted from $p\simeq 10^{-5}$ to $5\times 10^{-6}$, because
the inside Hubble time is larger than that outside:
$H_{\rm in}^{-1}(\ti)/H_{\rm out}^{-1}(\ti)=\sqrt{\rho_{\rm out}(\ti)/
\rho_{\rm in}(\ti)}=100$.  From
Eqn({\ref{eq:DeltHGR}}), the $1^{\rm st}$-crossing time is $\Delta t_c
\simeq 50/10(1+25/(2\times 10))=11.25$, in rough agreement
with the numerical collapse time of $\Delta t_c\simeq 8$.
Thus, the collapse time is found to be approximately equal
to the $1^{\rm st}$-crossing time---the the shock moves inward
at roughly the speed of light.
This is not surprising however, because the speed of sound is $.57c$ (III-C).

Figures 9b and 9c show the pressure as a function of $R~R(\ti,R_B)/R(t,R_B)$
for a general relativistic superhorizon-sized void with
$c^{-1}\RW/H^{-1}_{\rm out}(\ti)=25$, $\Gamma(\ti,R)=1$, $k^2=4$, $\rw=1$,
$\ep(\ti,R_B)=10^6$
and $\DR=.5$.
In addition,
$\alpha=10^{-6}$ and $R_{j_B}(t)=100.2,~125.1$, and $148.7$ for Figure 9b,
and $\alpha=10^{-10}$ and
$R_{j_B}(t)=100.2,~102.6,~{\rm and}~104.4$
for Figure 9c.
The numerical collapse times in Figures 9b and 9c are $\Delta t_c=1.3$ and
$.09$,
respectively, which are both smaller than the outside Hubble time.
(Note that $\Delta t_c=.09$ is $1/20^{th}$ the outside Hubble time).
Since $\Phi(\ti)$ equals
$31.6$ and $316$, respectively, the $1^{\rm st}$-crossing times from
Eqn({\ref{eq:DeltHGR}}) are $\Delta t_c\simeq 50/31.6(1+25/31.6)
\simeq 2.2$ and $50/316(1+25/316)=.17$, respectively.
The $1^{\rm st}$-crossing times are larger
than the numerical collapse times because
$\Phi_{\rm in}(t)$ increases during the collapse due to the dissipation
of energy at the shock.  Thus, the gain in entropy at the shock
only makes the collapse time shorter.  For example, in Figure 9c,
$\Phi_{\rm in}(t)$ increases from its original value of $\Phi_{\rm in}=316$
to $\Phi_{\rm in}\sim 450-500$ during the collapse.
Using $\Phi_{\rm in}=500$, we would predict $\Delta t_c\sim .1$, which is
roughly
correct.

We note however, that in Figure 9c (and to a lesser extent in Figure 9b),
only part of the void has been filled in at the collapse time $t_c\equiv\ti
+\Delta t_c$.
Thus, thermalization and homogenization has not been achieved by $t_c$.
We note from Figures 9a,b and c that as the initial relative energy density
inside
the void
decreases, the fraction of the void filled in at the collapse time decreases.
However, since the energy density inside the somewhat filled void
is still much less than $\rho_{\rm out}(t_c)$, ~$\Phi(t_c)$ inside the ``void''
is still much greater than one.  And because $\Gamma(t_c)$ is also larger than
one inside the ``void'', the distance light can travel is still greater inside
than outside the void.  Thus, although the void has not yet
homogenized, this
may only take an additional small amount of time.

Figures 10a and 10b show the pressure versus $R~R(\ti,R_B)/R(t,R_B)$
for a superhorizon-sized uncompensated and compensated void, respectively.
Here,
$\Gamma(\ti,R)=1$,
$\RW=500$, $\rw=10.$, $k^2=4$, $\ep(\ti,\rmx)=10^{6}$, $\DR=5.$,
and $\alpha=10^{-10}$.
(These voids are $250$ times the outside Hubble radius).
The pressure is shown at $\ti=1$ and $t=1.7$ for each void, and at
$t=2.1$ and $t=2.0$ for the uncompensated and compensated voids, respectively.
In addition, in Figure 10a, $R_{j_B}(t)=1002,~1288,~{\rm and}~1424$,
while in Figure 10b, $R_{j_B}(t)=1002,~1307,~{\rm and}~1418$.
The important point to note is that the shock reaches the origin at
$t\simeq 2.1$ for both voids.  This is due to the diffusion of particles into
the
void, and does
not depend on the compensatedness of the void because $\Phi_{\rm in}(\ti)$
depends only on $\rho_{\rm in}(\ti)/\rho_{\rm out}(\ti)$.
(This is similar to that for special relativistic
relativistic voids (VII-B)).
This roughly agrees with the predicted collapse time
$\Delta t_c=500/316(1+250/(2\times 316))\simeq 2.2$ from
Eqn({\ref{eq:DeltHGR}}).
Note also that at $t\simeq 2$, the difference between Figures 10a and 10b
is small.  For the compensated void, there is no perceptible outbound shock,
and the original
density ``bump'' in the void wall
is stretched and damped out.  In fact, the fluid initially in the
compensated void's wall is moving inward at $t\simeq 2.1$.

Up to this point, we have shown the evolution of
general relativistic uncompensated and compensated voids for the initial
velocity profile $U=U_{\rm GRAV}=\sqrt{2G_N M/R}$ (or $\Gamma(\ti,R)=1$), where
the initial velocity per shell just balances gravity.
If the wall initially has an outward peculiar velocity (as expected from
first-order inflation, for example),
then $U(\ti,\RW)>U_{\rm GRAV}(\ti,\RW)$ (or $\Gamma(\ti,\RW)>1$), as discussed
in II-D.
In Figure 11, we show $4\pi p$, $M$, $\Gamma$ and $\Phi$ versus
$R~R(\ti,R_B)/R(t,R_B)$
for a compensated superhorizon-sized void
with $U/c=R\sqrt{8\pi G_N\rho/3}$ (Eqn({\ref{eq:velbig}})),
$\RW=500$, $\rw=10.$, $k^2=4$, $\ep(\ti,\rmx)=10^{6}$,
$\DR=2.5$, $\f=.005$ and $\alpha=10^{-10}$.  The voids are superhorizon-sized:
$\RW/H(\ti,\rmx)^{-1}=250$.
In addition, $R_{j_B}(t)=601.2,~783.9,~{\rm and}~911.8$,
and we show the configurations at $\ti$ and at times
$t=1.7$ and $t=2.3$.
The void wall is initially moving outward with a very large
peculiar velocity, since $\Gamma(\ti,\RW)\simeq 900$.

We find the very interesting result that even though the wall has
a large outward peculiar velocity, the inner part of the void still
collapses.  This is because the fluid near the base of the void wall
gets accelerated into the void right away, pulling adjacent fluid with it.
It is true however, that at $t\sim 2$
the density profile looks different than that from
Figure 10b.  This is because the fluid
in the void wall takes more time to lose
its outward velocity.
Since the numerical collapse time is $\Delta t_c=1.3$ from Figure 11, we find
that the extra time taken
for the void
to collapse is approximately $1.3-1.1\simeq .2$, which is still less
than the outside Hubble time.
(This follows qualitatively from our discussion in Section II-D, where it was
argued
that $\Gamma$ would decrease very quickly at the inner edge of the wall,
since $\dG\propto -\Gamma^2 p'$).
These new results show that the collapse time of a superhorizon-sized
void with a large outward peculiar wall velocity can be of order
the $1^{\rm st}$-crossing time.  Since the minimum thermalization
and homogenization time is the $1^{\rm st}$-crossing time,
the time for thermalization and
homogenization of this void may be short.

In Figure 12, we show the void radius versus
cosmic time for a void with initial size $\RW=c10^{23}\Hm$ .  If the
temperature
outside the void initially is $T_{\rm out}(\ti)=10^{14}$GeV,
then the initial time is $\ti=10^{-33}$ seconds.$^{\ref{TEU}}$
And because recombination occurs at $t_{\rm rec}\simeq 10^{12}$ seconds,
$t_{\rm rec}/\ti=10^{45}$, which is near where the dashed lines intersect
at the top of Figure 12.  Therefore, the plot consists
the radiation dominated epoch in the early universe, where
the scale factor outside the void is $a(t)=a(\ti)\sqrt{t/\ti}$.
Because $R(\ti,R_B)/R(t,R_B)=\sqrt{\ti/t}$, any comoving point
(defined as $R\propto\sqrt{t/\ti}$, not $r=const$)
is a vertical line in this plot.
The dashed line labeled $r_{\rm CM}$ shows the void size if it were comoving
with spacetime, as previously suggested (see IV).  The Hubble radius outside
the void is
$cH^{-1}_{\rm out}\equiv R_{\rm HOR}=2ct$, and is shown as the dashed line
labeled by $r_{\rm HOR}$.  As can be seen, $r_{\rm HOR}$ and $r_{\rm CM}$
intersect at $t/\ti\simeq 10^{46}$ (or $10^{-11}$ seconds), shortly after
recombination.
(This also follows from Eqn({\ref{eq:conf}}), since
$\Delta t_c\simeq \Hm(c^{-1}\RW/\Hm)^2\simeq 10^{46}$).
If a superhorizon-sized void were to conformally expand with spacetime, then
this void would just barely be around to distort the microwave background
at recombination.$^{\ref{TurWeiWid}}$

The dash-dot lines show the position of the $1^{\rm st}$-crossing
photons (i.e. the inner void wall) from
Eqn({\ref{eq:radwall}}).
We show
$\Phi_{\rm in}(\ti)=5\times 10^{11}$, $10^{15}$ and $10^{20}$
(since we require $\Phi_{\rm in}(\ti)\ga \sqrt{10^{23}}$
(from Eqn({\ref{eq:rwvalid}}))), with
the predicted $1^{\rm st}$-crossing times of
$t_c/\ti\simeq\Delta t_c/\ti=4\times 10^{22}$, $10^{16}$
and $10^6$ respectively.  If $\Delta t_c/\Hm>1$,
the radius of the inner edge of the wall is constant ($R\simeq const$) until
time $t\simeq t_c\simeq \ti(\Phi^{-1}_{\rm in}(\ti)\RW/\Hm)^2$,
at which point $R$ rapidly decreases to zero (see Eqn({\ref{eq:radwall}})).
If the reheat temperature
is $T_{\rm RH}=T_{\rm out}(\ti)=10^{14}$GeV, then
the void with $\Phi_{\rm in}(\ti)=T_{\rm out}(\ti)/T_{\rm in}(\ti)=10^{20}$
may thermalize by $t\simeq 10^{-27}$
seconds (or at $T\simeq T_{\rm RH}\sqrt{\ti/t}\simeq 10^{11}$GeV),
far before recombination or nucleosynthesis.  (Note that this value of
$\Phi_{\rm in}(\ti)$
corresponds to an initial void temperature of
$T_{\rm in}(\ti)=T_{\rm RH}\Phi_{\rm in}(\ti)^{-1}=10^{-6}{\rm GeV}=1$keV).

In conclusion, we have seen that a compensated or uncompensated
superhorizon-sized
void collapses as fluid in the wall diffuses into the void at the speed of
light.
Thus, the collapse time is found to be approximately equal to the
$1^{\rm st}$-crossing time
calculated in Section IV, which is {\it much} smaller than previously
thought.

\centerline{\bf{VIII. Discussion}}
\setcounter{section}{8}
\setcounter{equation}{0}

In this paper, the evolution of a superhorizon-sized void
embedded in a Friedmann-Robertson-Walker (FRW) universe was studied
by numerically evolving the spherically symmetric general relativistic
equations
in the Lagrangian gauge and synchronous coordinates.
(A superhorizon-sized void is a void larger than the Hubble radius outside the
void:
$c^{-1}\RW/$ $\Hm>1$, where $\ti$ is the initial cosmic time,
$c H^{-1}_{\rm out}(\ti)$ is the Hubble radius outside the void
and $\RW$ is the radius of the void).
The particles are
assumed to be in local thermal equilibrium so that they can be described
as a fluid, and the perfect fluid equation of state is chosen.
The inside of the void is initially chosen to be homogeneous and
non-expanding.
We are particularly interested in the evolution of voids with
$T/\mu\gg 1$, where $T$ is the local temperature
and $\mu$ is the mass of a fluid particle.  This corresponds to a
radiation-dominated period in the early universe.
We find that for $p=\rho/3$, general relativistic voids
collapse via a shock propagating inward from the void wall.
The shock is formed from
the steep pressure gradient at the inner edge of the wall.
It moves at approximately the speed of light; this
is not surprising, since the speed of sound is $.57c$.
Thus the void collapse time (the time taken for the shock
to reach the origin) roughly equals the photon
$1^{\rm st}$-crossing time (the time taken for a photon initially at the inner
wall edge to reach the origin).  At the time this shock reaches the origin,
much of the fluid in the original wall area is moving toward the origin
behind the shock.
The energy density of this fluid (i.e. the fluid behind the shock) is less
than that outside the
void.  At the collapse time then, only part of the void has been filled in with
fluid.  In particular, as the initial energy density inside the void
decreases, the fraction of the void filled in at the collapse time decreases.

Because the shock moves inward at roughly the speed of light, we can calculate
the approximate collapse time.
For $p=\rho/3$, the $1^{\rm st}$-crossing time is
\be
\label{eq:cross}
\Delta t_c=c^{-1}\RW \left(\frac{\rho_{\rm in}(\ti)}{\rho_{\rm out}(\ti)}
\right)^{1/4}
\left[1+\frac{c^{-1}\RW}{2H^{-1}_{\rm out}(\ti)}
\left(\frac{\rho_{\rm in}(\ti)}{\rho_{\rm out}(\ti)}\right)^{1/4}\right]
\ee
for $\RW/\Hm\la\sqrt{\rho_{\rm out}(\ti)/\rho_{\rm in}(\ti)}$, where
$\rho_{\rm in}(\ti)$ and $\rho_{\rm out}(\ti)$ are the initial
fluid energy densities inside and outside
the void, respectively, and
$\rho=T^4$.
If $c^{-1}\RW/H^{-1}_{\rm out}(\ti)<(\rho_{\rm out}(\ti)/\rho_{\rm in}(\ti))^
{1/4}$,
then $\Delta t_c/\Hm\la 1$; the $1^{\rm st}$-crossing time
is less than or comparable to the initial Hubble time outside the void.
In fact, as
the density inside the void approaches zero, the collapse time
goes to zero!

It may seem contradictory that the $1^{\rm st}$-crossing time for
a {\it superhorizon-sized} relativistic void can be less than the outside
Hubble
time (i.e. or that light travels comparatively farther inside a void than
outside).
There are several points to make about this.  First,
if we choose Eulerian instead of Lagrangian synchronous coordinates
to evolve special relativistic voids ($G_N=0$),
then the $1^{\rm st}$-crossing time is not fast.  This is because spacetime
is sliced differently in time inside the void.  Since a similar
$1^{\rm st}$-crossing time is obtained for general relativistic voids,
it might be argued that the quick collapse time is a figment of
the Lagrangian gauge used.  However because the FRW metric
has a coordinate singularity at the Hubble radius when transformed
to Eulerian synchronous
coordinates, superhorizon-sized voids cannot be evolved in these
coordinates.  Any prior intuition from special relativistic voids
in these coordinates therefore, must be carefully applied.

Second, calling a void ``superhorizon-sized'' is deceptive.
Although the $1^{\rm st}$-crossing time for a superhorizon-sized
{\it perturbation} in the FRW universe is greater than the outside Hubble time,
this does not imply that the same is true for a superhorizon-sized void.  This
is because the void is not a small perturbation in general (i.e. $(\rho_{\rm
out}-
\rho_{\rm in})/\rho_{\rm out}\simeq 1$).
If the void's size is defined to be the spacelike circumferential radius
relative to the {\it outside} Hubble radius, then the void is
superhorizon-sized
if $R_{\rm wall}> cH^{-1}_{\rm out}$.
However, if the voids's size is defined to be its radius
relative to the
Hubble radius {\it inside} the void, then its size is smaller since
$c^{-1}\RW/H^{-1}_{\rm in}(\ti)=c^{-1}\RW/H^{-1}_{\rm out}(\ti)
\sqrt{\rho_{\rm in}(\ti)/\rho_{\rm out}(\ti)}$; it is {\it subhorizon-sized}
if $c^{-1}\RW/\Hm<\sqrt{\rho_{\rm out}(\ti)/\rho_{\rm in}(\ti)~}$ !
(We call this latter size the ``radial size'', because it is the time
taken for a photon to cross a static void multiplied by $c$).
This is because
the Hubble radius is much larger inside than outside the void.
Since the void is an underdense
region, we expect the opposite relative size problem to occur for overdense
regions where the gravitational potential is large, not small.
As an example, a black hole's circumferential size
is small (or at least finite), while its ``radial size'' is infinite.

The original motivation
for studying this problem was to determine the evolution of
voids formed from first-order (e.g. extended) inflation.
These voids are compensated in energy density, and the walls
initially have large outward peculiar velocities.$^{\ref{Vasha}}$
Previous authors estimated the minimum homogenization time by
determining the $1^{\rm st}$-crossing time.  They found it
to be$^{{\ref{Wein}},{\ref{TurWeiWid}},{\ref{LiddWan}}}$
\be
\label{eq:erick}
\Delta t_c\simeq H^{-1}_{\rm out}(\ti) \left(\frac{c^{-1}\RW}{\Hm}\right)^2.
\ee
This estimate was based on the assumption that
a superhorizon-sized void would conformally expand with spacetime.
The present work throws considerable doubt on this assumption
and its implications (i.e. a long thermalization time for superhorizon-sized
voids) for several reasons.
First, the $1^{\rm st}$-crossing time is much shorter than this
estimate.  For $\Delta t_c/\Hm>1$, this estimate is off by the factor
$\sqrt{\rho_{\rm in}(\ti)/\rho_{\rm out}(\ti)}$ (see Eqn({\ref{eq:cross}})).
Since the minimum thermalization time is the
$1^{\rm st}$-crossing time, this implies that thermalization and
homogenization may happen much quicker than suggested.
This would have profound effects upon our understanding of the
evolution of superhorizon-sized voids in the early universe.
Second, the inner edge of the void collapses (not expands)
in roughly the $1^{\rm st}$-crossing time, even (apparently)
for voids with large outward peculiar velocities.
This collapse occurs because the
wall decelerates from the positive
wall pressure ($\dU\propto -p'<0$).  If the wall pressure were
negative instead (e.g. the wall pressure of a vacuum bubble), the wall
would accelerate outward.
Thus, fluid from at least part of the wall rushes into
the void, partially ``filling it'' at the $1^{\rm st}$-crossing time.
Because the light travel
distance is still large there, the void may still fill up in a relatively short
amount of time.
Third, the void wall can cease expanding by many mechanisms.
Because fluid rushes into the void, large
velocity gradients are formed in the original wall area
which ``pull'' on the wall and can
slow it down.  In addition, when the peculiar wall velocity is much larger
than the gravitational velocity, the deceleration of the wall is proportional
to the velocity squared which can be enormous.
Finally, an outgoing wave will be damped and slowed down in any case,
by virtue of the volume effect caused by the spherical geometry.
Thus even if the wall initially expands outward conformally
(or faster), it may stop doing so rather quickly,
as the simulations appear to suggest.
In any case, even if the outer part of the wall {\it can} carry out some mass
for a long period of time, the interior region will continue to fill in and
homogenize.
Since overdensities will most likely form near the origin after collapse,
perturbations, small waves and/or overdensities
will most likely be around at recombination.  These perturbations could have
interesting amplitudes, and
therefore could alter the standard Harrison-Zeldovich density spectrum.
In any case, because the qualitative void evolution picture
previously suggested is at least partially
incorrect, it is reasonable to expect the estimated thermalization
time to be erroneous; the ``big-bubble problem'' might not
be a problem after all.

It is important to point out that
in order not to distort the microwave background, the thermalization time only
needs to decrease somewhat from the estimate given by Eqn({\ref{eq:erick}}).
The largest superhorizon-sized void from minimalist inflation is roughly
$c^{-1}$ $\RW/\Hm\simeq 10^{27}$. $^{\ref{TurWeiWid}}$   If the initial time
after reheat is $\ti=10^{-33}$ seconds ($T_{\rm RH}=10^{14}$GeV),
then at recombination ($t_{\rm rec}=10^{12}$ seconds),
this void would not have thermalized according to Eqn({\ref{eq:erick}).
Using these initial conditions with Eqn({\ref{eq:erick}}), it is easy to see
that voids for which $c^{-1}\RW/\Hm\ga 10^{23}$ were traditionally
troublesome.
However, since $10^{23}$ is such an enormous number, a very
slight change in the functional form for the thermalization time,
$\Delta t_H $, can cause thermalization to occur before
recombination.  For instance, if we suppose that the thermalization
time can be written as
$\Delta t_H\simeq c^{-1}\Hm(\RW/\Hm)^p$, then an acceptable range
for $p$ is $p\la 1.6$, which is not dramatically different from $2$.  The point
is that the ``big-bubble problem''
can only be resolved by {\it accurate} knowledge of the thermalization
time, because estimates off by a somewhat small amount can lead
to erroneous conclusions.

Of the many questions which remain unanswered, the
most important is to determine what happens to a superhorizon-sized
void after it collapses.  If it thermalizes and homogenizes,
at what time does this happen?  As mentioned previously,
at the time of $1^{\rm st}$-crossing,
the energy density within the former void region is greater than
$\rho_{\rm in}(\ti)$, but is still less than $\rho_{\rm out}(t)$.  In addition,
the bulk motion of much of the original wall fluid is inward so that spacetime
is collapsing
and not expanding there;
thermalization and homogenization has not occurred.
However, since the ``void'' is still relatively underdense at this time,
the light travel distance will be comparatively large,
thereby allowing for the possibility of relatively fast
thermalization and homogenization.
We have found that special relativistic voids eventually nearly homogenize
after
creating an {\rm overdense} region at the origin.  In this case, all but the
fluid near the origin becomes roughly homogeneous fairly quickly.  The fluid
near the origin however, is overdense for a relatively long time.
If general relativistic voids behave similarly, then after collapsing, an
{\it overdense} region would form at the origin.  This would
take a relatively long amount of time to diffuse away, and would eventually
result in a nearly homogeneous and isotropic universe.
In any case, since this overdense region would form very quickly,
it is unlikely that empty voids would be around any time
near recombination, as previously
suggested.$^{{\ref{La}},{\ref{TurWeiWid}},{\ref{LiddWan}}}$
Instead, overdensities, perturbations and small waves with strange fluid
velocities may be.
Future work will study the consequent evolution of a general relativistic
void after collapse,$^{\ref{Vasha2}}$
as well as the evolution of a superhorizon-sized void from
first-order (e.g. extended) inflation.$^{\ref{Vasha}}$

\centerline{\bf{Acknowledgements}}

The author wishes to thank her thesis advisor, Rocky Kolb, for suggesting
this problem and for helpful discussions.
She also wishes to thank Andrea Malagoli, Rick Watkins,
Sidney Bludman, Michael Turner, Marc-Mordecai Maclow, Paul Steinhardt
and Tsvi Piran
for interesting and useful discussions.
She would also like to thank Stephen Arendt for useful discussions
and help with the manuscript.
This work was supported in part by NASA under Grant NAGW-2381 at Fermilab and
by the DOE at Chicago and Fermilab.

\vspace{.1in}

\centerline{\bf{Appendix A:  Transformation of the FRW metric to Eulerian
Coordinates}}
\@addtoreset{equation}{section}
\def\ksection{\arabic{section}}
\def\theequation{A.\arabic{equation}}
\setcounter{equation}{0}

As discussed in Section IIA., the metric in Eq.~(\ref{eq:mostgen}) can be
subjected to
transformations of the type $t=f_1(t',r') ~{\rm and~} r=f_2(t',r')$ without
altering its spherical symmetry.  If we set $r^2=k(t,r)$,
then we can rewrite Eq.~(\ref{eq:mostgen}) as
\ba
\label{eq:metric0}
ds^2&=&-(N^2-S^2)dt^2+2Sdtdr+h^2dr^2+r^2 d\Omega^2,
\ea
where we have introduced
the lapse and shift functions as $N(t,r)$ and $S(t,r)$, respectively.
This is the Eulerian gauge with synchronous
coordinates if we choose $S=0$.
As we shall
see, when the FRW spatially flat metric is transformed in this manner,
the resulting metric has a coordinate singularity at the
horizon.
Thus, the study of superhorizon-sized voids is not possible in
these coordinates.

Consider the FRW $k=0$ metric given by Eqn~({\ref{eq:FRW}}).
Letting $\rtw=ra(t)$, we can rewrite this as
\be
\label{eq:metric2}
ds^2=-\left(1-\rtw^2\left(\frac{\dot{a}}{a}\right)^2\right)dt^2-2\rtw
\frac{\dot{a}}{a}d\rtw dt
+d\rtw^2+\rtw^2 d\Omega^2,
\ee
where we set $c=1$ everywhere in this appendix.  This is the Eulerian
asynchronous form of the FRW metric.
We notice that this takes the form of Eq.~(${\ref{eq:metric0}}$)
with $N=1$, $S=\rtw \dot{a}/a$ and $h=1$.
Since $a(t)\propto t^\p$, where $\p=1/2~{\rm and}~2/3$
in radiation ($p=\rho/3$)
and matter ($p=0$) dominated
universes respectively, we can eliminate the $d\rtw dt$ term by
defining
\be
\label{eq:tgtw}
t=\p\rtw/g(\ttw,\rtw)
\ee
and ensuring that the condition
\be
\frac{\partial g}{\partial\rtw}=g~\frac{\p+(1-\p)g^2}{\p\rtw(1-g^2)}
\ee
hold.  Defining $\kappa\equiv (1-\p)/\p$ and
$\alp\equiv 1/[2(1-\p)]$, this last expression can be integrated
to obtain
\be
\label{eq:htw}
h(\ttw)\rtw=g/(1+\kappa g^2)^\alp,
\ee
where $h(\ttw)$ is an arbitrary (integration) function of $\ttw$ only.
The metric then becomes
\be
\label{eq:newmet}
ds^2=-\left(\frac{h\rtw}{g}\right)^4 \frac{(1+\kappa g^2)^
{2(\alp+1)}}{1-g^2}d\ttw^2+\frac{1}{1-g^2}d\rtw^2+\rtw^2 d\Omega^2.
\ee
This metric has a coordinate singularity at $g=1$, which from
Eqn({\ref{eq:tgtw}})
occurs at $r=t/a(t)~\p^{-1}$.  But this is the comoving radius of the
Hubble radius ($r_{\rm HOR}$), since $H^{-1}=t/\p=r_{\rm HOR} a(t)$.
Therefore, we have shown that
for the FRW models expressed in Eulerian, synchronous coordinates, a
coordinate singularity occurs at the Hubble radius.

As an interesting
example we take $\p=1/2$,
which corresponds to
radiation-domination ($p=\rho/3$).
In addition, we choose
$h\equiv \p/\ttw$.
Using Eqs.~(${\ref{eq:htw}}$) and
(${\ref{eq:tgtw}}$) together with
$\rtw=r~a(t)=r~a(\ti)(t/\ti)^{1/2}$,
we find that $\alpha=\kappa=1$, $\ttw=t+(a(\ti) r)^2/(4\ti)$ and
$g=(1-\sqrt{1-4(\rtw/\ttw)^2})/(\rtw/\ttw)$.
In addition, the metric simplifies to
$ds^2=-(1-g^2)^{-1}d\ttw^2+(1-g^2)^{-1}d\rtw^2+\rtw^2 d\Omega^2$.
Note that even though the energy density is spatially constant on
a hypersurface $t=constant$, on a hypersurface of constant
$\widetilde t$, ~$\rho\propto t^{-2}=4g^2\widetilde r^{-2}
=4\widetilde t^2(1-\sqrt{1-4(\widetilde r/\widetilde t)^2}~)^2/\widetilde r^4$
which is {\it not} spatially constant.

\centerline{\bf{Appendix B:  Differencing of the Equations}}
\@addtoreset{equation}{section}
\def\ksection{\arabic{section}}
\def\theequation{B.\arabic{equation}}
\setcounter{equation}{0}

In this appendix, we list the expressions as they are differenced in the
code.  We use the shorthand notation
$\Delta G\equiv G_{i+1}^j-G_i^j$ or
$\Delta G\equiv G_{i}^j-G_{i-1}^j$ for forward or backward differencing,
respectively.  The equations
involving spatial derivatives are
\ba
\label{eq:diffshc}
\dU&=&-\Phi\left(\frac{G_NM}{R^2}+\frac{4\pi G_N(p+Q)R}{c^2}\right)-
3\frac{\Gamma\Phi R^2}{w \Delta(r^3)}
4\pi \Delta(p+Q) \nonumber\\
\dn&=&-\frac{n\Phi\Delta(R^2U)}{R^2\Delta R}\nonumber\\
\dep&=&-\frac{4\pi\Phi (p+Q)\Delta(R^2U)}{\Gamma r^2\Delta r},
\ea
We note that
there are many other schemes which could potentially be used.
(For example, one  could write $\Delta r^3/3$ instead of $r^2 \Delta r$
in the expression for $\dep$~).
Although these variations can
produce and propagate the shock and rarefaction waves, they can not
handle the bounce of the shock at the
origin.  The only set of difference equations which we have
found to do this are given above.
For the Tolman-Bondi
pressureless dust models (Section V), it is found that
more accurate solutions are obtained near the origin when differencing the
$\dn$
equation as follows:
$\dn=-({n^2\Phi\Delta(R^2U)})/({\Gamma r^2\Delta r})$.  The
Tolman-Bondi figures shown
in this paper, however, use the difference equations as written
in Eqns({\ref{eq:diffshc}}).

\centerline{\bf{Appendix C:  General Relativistic Jump Conditions}}
\@addtoreset{equation}{section}
\def\ksection{\arabic{section}}
\def\theequation{C.\arabic{equation}}
\setcounter{equation}{0}

In this appendix, we derive the jump conditions for general relativistic
shocks.  This closely follows the derivation by
May and White (1965)$^{\ref{MayWhite}}$.
These conditions will then be used to derive a more
general artificial viscosity expression.
In addition, the shock jump conditions for ultra special relativistic
shocks are derived.

Let the variables $a$ and $b$ represent labels for comoving observers
ahead and
behind the shock, respectively.  If each observer measures the invariant
interval separating the same two events on the world surface of a shock,
using Eqn({\ref{eq:metric}}), we obtain
\be
\label{eq:shmet}
[ds^2]=[-c^2\Phi^2 dt^2+\Lambda^2dr^2+R^2d\Omega^2]=0,
\ee
where $[G]\equiv G_a-G_b$.  Because the shocks
are radial, we can choose the two events to have
$dr=dt=0$ but $d\Omega\neq 0$.
Therefore, $R$ is continuous across the shock: $[R]=0$.
Now consider two events with $dr\neq 0$ and $dt\neq 0$.  Then from
Eqn({\ref{eq:shmet}}),
\be
\label{eq:phi2}
[c^2\Phi^2-M_s^2/n^2]=0,
\ee
where $r_s$ is the position of the shock in comoving coordinates,
$dr_s/dt$ is the shock ``speed'', and
$M_s\equiv {f ~(dr_s/dt)}/({4\pi R^2})$ from
Eqns({\ref{eq:defn}}) and ({\ref{eq:Gamlam}}).
This is the first of the jump condition equations.

It can be shown that the jump conditions
for the Schwarzschild metric (but {\it not} for the comoving metric)
are $^{{\ref{Synge}},{\ref{MayWhite}}}$
\be
\label{eq:cont}
[{T_\nu}^\mu\partial g/\partial x^\mu]=0,
\ee
where $g$ is the equation for the world surface of the shock.
Therefore we must first find the jump
conditions in the Schwarzschild frame and then transform back
to the comoving frame.
The Schwarzschild metric is
$ds^2=-c^2 A^2 dT^2+B^2dR^2+R^2d\Omega^2$,
where $R$ is now the ``Eulerian'' coordinate radius.  Since $R$ is
a coordinate, $[R]=0$, and using the same argument as
above for the two observers,
\be
\label{eq:new}
[c^2A^2-B^2S^2]=0,
\ee
where $S$ is the shock ``speed''
in this frame: $cS\equiv dR_s/dT$.  For the Schwarzschild metric, the
solution for the metrics functions are well known: $B^2=1-2MG/(Rc^2)=A^{-2}$
where $M(R,T)=4\pi c^{-2}\int_0^R\rho R^2dR$.
As long as $\rho$
is not infinite in the shock,
the mass is continuous across the shock, $[M]=0$.
Then $[B]=[A]=0$.  And using Eqn~({\ref{eq:new}}),
we find that $[S]=0$.  We conclude that
in the Schwarzschild frame, the metric components and the shock ``speed''
are continuous across the shock: $[A]=[B]=[R]=[S]=0$.

We now derive the jump conditions in the Schwarzschild frame.
If a shock is located at position $R_s(T)$ at time $T$, the equation for the
world surface of a shock is
$g=R_s(T)-R=0$.
In addition, the perfect fluid stress-energy tensor in the Schwarzschild
frame\footnote{We
denote quantities in the Schwarzschild frame by primes.} is
\be
{T'_\mu}^\nu=c^{-2}(\rho+p)g'_{\mu\lambda}u'^\mu u'^\lambda+pg'_{\mu\lambda}
g'^{\nu\lambda}=
nwg'_{\mu\lambda}u'^\mu u'^\lambda+pg'_{\mu\lambda}g'^{\nu\lambda},
\ee
and the conservation of mass equation is $[nu'^\nu \partial g/\partial x^\nu]=
0$.~$^{{\ref{Synge}},{\ref{MayWhite}}}$
Using Eqns~({\ref{eq:cont}}), the junction conditions become
\ba
\label{eq:eqen}
\left[T_T^T
S-T_T^R\right]=c^2\left[(-A^2(u'^T)^2nw+p)S+nwA^2u'^Tu'^R\right]&=&0\\
\label{eq:eqmom}
\left[T_R^T S-T_R^R\right]=\left[SnwB^2u'^Tu'^R-(nwB^2(u'^R)^2+p)\right]&=&0\\
\label{eq:eqma}
\left[nSu'^T-nu'^R\right]&=&0.
\ea

We would like to express these equations in terms of
the comoving metric functions.  First,
we can rewrite the shock velocity in the comoving frame
as $cS=(R_t+R_r \rs)/(T_t+T_r\rs)$.  In addition, we can
relate the metric functions from the comoving frame to those in the
Schwarzschild frame by
$g'^{\mu\nu}=({\partial x'^\mu}/{\partial x^\sigma})~
({\partial x'^\nu}/{\partial x^\lambda})~g^{\sigma\lambda}$.
We then obtain $(cA)^{-2}=(c\Phi)^{-2}T_t^2-\Lambda^{-2}T_r^2$,
 $B^{-2}=-(c\Phi)^{-2}R_t^2+\Lambda^{-2}R_r^2$ and $(c\Phi)^{-2}T_tR_t=
\Lambda^{-2}T_rR_r$.
In addition, the ``mass'' $M(r,\t)$ contained within comoving radius $r$ is
continuous across the shock,
since $[M]=[nR^2(1+\ep/c^2)]dR~^{~~~\Rightarrow}_{dR\rightarrow 0} ~0$.
Using Eqn~({\ref{eq:defGamma}}), we find that $[\Gamma^2-U^2/c^2]=0$.

We can use these relations to calculate the 4-velocity as
measured in the Schwarzschild frame.
Since the 4-velocity in the comoving frame is $u^\mu=(-c\Phi^{-1},0,0,0)$,
and the velocity transforms as
$u'^\lambda=(\partial x'^\lambda/\partial x^\sigma)~ u^\sigma$,
$u'^\lambda=-\Phi^{-1}(c\dot{T},\dot{R},0,0)$,
where $\dot{T}=\partial T/\partial t$ and $\dot{R}=\partial R/\partial t$.
Since $U\equiv \dot{R}/\Phi$,
and using the fact that the 4-velocity is
normalized to $u'^\lambda u'_\lambda=-c^2=c^2A^2(u'^T)^2-B^2(u'^R)^2$, we can
rewrite the $4$-velocity as
$u'^\lambda=-(A^{-1}\sqrt{B^2U^2+c^2},U,0,0 )$.
In the special relativistic limit ($G_N=0$), $A=1~{\rm and~}B=1$ and this
becomes
\be
\label{eq:Gaminterp}
u'^\lambda = -((U^2+c^2)^{1/2},U,0,0)= -(c\Gamma,\Gamma {v},0,0),
\ee
where we have defined ${v}\equiv U/\Gamma$ so that
$\Gamma=(\sqrt{1-(v/c)^2})^{-1}$.
In the special relativistic limit, then, ${v}$ is the fluid
particle's radial velocity, $\Gamma$ is usual gamma-factor (i.e. energy per
particle mass),
and $U$ is the particle momentum per particle mass.

We can now rewrite Eqn~({\ref{eq:eqen}})-({\ref{eq:eqma}})
and $[S]=0$ in terms of the
comoving metric functions using the fact that
$T_r/T_t=c^{-2}U\Lambda/(\Gamma\Phi)$ and $[\rs]=0$. We obtain
\ba
\label{eq:ma}
[U M_s/(nc^2)+\Phi\Gamma]&=&0\\
\label{eq:en}
[c^2M_s\Gamma(1+\ep/c^2)-pU\Phi]&=&0\\
\label{eq:mom}
[M_s U(1+\ep/c^2)-p\Gamma\Phi]&=&0\\
\label{eq:shvel}
[M_s\Gamma/n+\Phi U]&=&0.
\ea
Eqn~({\ref{eq:phi2}}) and Eqns~({\ref{eq:ma}})-({\ref{eq:shvel}})
make up the required shock conditions.
One of them however, is redundant.
It can be shown that in the nonrelativistic limit,
the above conditions reduce to the
Lagrangian shock jump conditions given in Ref~${\ref{CourFrei}}$.

The case considered by May and White (1967)
is the penetration of a shock into a non-relativistic medium.
We set $G_N=0$ and $\ep_a=p_a=U_a=0$ and take
$\ep=p/[(\gamma-1)n]$.
Using Eqns~({\ref{eq:en}}) and ({\ref{eq:mom}}) with the
fact that $U_b^2=\Gamma_b^2-1$, the energy behind the shock
is found to be $\ep_b=\Gamma_b-1$.  Using Eqns~({\ref{eq:mom}})
and ({\ref{eq:shvel}}), we eliminate $\Phi_b$ and
find that $\eta\equiv n_b/n_a=(1+\Gamma_b \gamma)/(\gamma-1)$.  From
Eqn~({\ref{eq:mom}}), we find that
\be
\label{eq:MPhi}
M={(\gamma-1)n_b\ep_b}\Phi_b/{U_b},
\ee
and plugging this into Eqn~({\ref{eq:phi2}}),
we determine that $\Phi_b=1/(1+\gamma(\Gamma_b-1))$.
Finally, we plug this into Eqn~({\ref{eq:MPhi}}) and find
$M_s=cn_a~\sqrt{(\Gamma_b-1)/(\Gamma_b+1)}~(1+\Gamma_b\gamma)/
[1+\gamma(\Gamma_b-1)]$.
(It turns out that Eqn~({\ref{eq:ma}}) gives no new information).
Thus, all quantities behind the shock can be expressed in terms of
$\Gamma_b$, the ``strength'' of the shock.

How does this affect the artificial viscosity needed to smooth the
shock?
Von-Neumann's viscosity$^{\ref{VonNRich}}$ is
$Q=k^2n(U')^2dr^2$ for $U'<0$ and $0$ otherwise, with
the shock being spread over roughly $k^2$ grid points.
Over the $k^2$ points, $Q\sim n(U_b-U_a)^2=nU_b^2=c^2n(\Gamma_b^2-1)=
c^2n(\Gamma_b-1)(\Gamma_b+1).$
Using the fact that $\ep_b/c^2=\Gamma_b-1$,
$Q\sim n\ep_b\Gamma_b\sim p\Gamma$ which is greater than the pressure
$p$ by the factor of $\Gamma\geq 1$.  Since
$Q$ should be of order the pressure in the shock region, we divide
by $\Gamma$:
$Q=k^2n(U')^2dr^2/\Gamma$ for $U'<0$ and $Q=0$ otherwise.
In the nonrelativistic limit, $\Gamma=1$ and
Von-Neumann's artificial viscosity is obtained.

In this paper, we are more
interested in the case where
the fluid on both sides of the shock is ultrarelativistic.
We first set $G_N=0$ and $U_a=0$.
Then Eqn({\ref{eq:phi2}}) and Eqns({\ref{eq:ma}})-({\ref{eq:shvel}})
can be manipulated to yield
\ba
\label{eq:Ub}
U_b&=&\sqrt{\Gamma_b^2-1}\\
\label{eq:M}
M&=&cn_a\eta{\sqrt{\Gamma_b^2-1}~\Phi_a}~/~({\Gamma_b\eta-1})\\
\label{eq:eta}
\eta&=&{\Gamma_b(1+\gamma\ep_b/c^2)-(1+\ep_a/c^2)}~/~[{\ep_b/c^2(\gamma-1)}]\\
\label{eq:epb}
\ep_b/c^2&=&-1+\Gamma_b(1+\ep_a/c^2)+(\gamma-1)(\Gamma_b-\eta^{-1})\ep_a/c^2\\
\label{eq:phib}
\Phi_b&=&{|\eta-\Gamma_b|}~\Phi_a~/~({\Gamma_b\eta-1}).
\ea
(It can be shown that these equations reduce correctly
in the $\ep_a\rightarrow 0$ limit).
Now, if we set $\gamma=4/3$
and take the ultrarelativistic limit, $\ep_b/c^2>\ep_a/c^2>>1$,
Eqn~({\ref{eq:eta}}) becomes $\eta= 4\Gamma_b\{1-3\ep_a/(4\Gamma_b\ep_b)\}
\simeq 4\Gamma_b$.
Then the equations can be approximately solved to give
\ba
\ep_b&\simeq &{4}\ep_a\Gamma_b(1-(4\Gamma_b)^{-2})~/~{3},
\ea
which to lowest order is $\ep_b\simeq 4\ep_a\Gamma_b/3$.
Therefore, $\ep_b$ depends not only on $\Gamma_b$ but also on $\ep_a$.
We generalize the artificial viscosity to
\ba
\label{eq:artvis}
Q&=&k^2n\left(1+{\ep}/~({\Gamma c^2})\right){(U')^2dr^2}/~{\Gamma}~~~{\rm
for~U'<0}\nonumber\\
Q&=&0~~~~~~~~~~~~~~~~~~~~~~~~~~~~~{\rm otherwise}.
\ea
For a strong shock then, $Q\sim n\ep U^2/(\Gamma^2 c^2)\sim
n\ep\sim p$, as desired.
And when $c^2\rightarrow\infty$, this becomes VonNeumann's original
expression.
Note that because this was derived in the limit of strong shocks,
it does not work as well for weaker special or general relativistic
shocks.$^{\ref{Vasha2}}$

We now write
down the exact solution of Eqns~({\ref{eq:Ub}})-({\ref{eq:phib}}) for
conditions behind the shock in terms of $\Gamma_b$ only and quantities
in front of the shock.  Because $\rho=n\ep$ in this limit, we can rewrite
Eqns~({\ref{eq:eta}}) and ({\ref{eq:epb}}) as $\eta=4\rho_b\Gamma_b/
(3\rho_a+\rho_b)$ and
$\eta=(3\rho_b+\rho_a)/(4\rho_a\Gamma_b)$ respectively.  Setting them
equal, we can solve for $\rho_b$.
Then, all others quantities are determined.  They are
\ba
\label{eq:Ur}
U_b&=&\sqrt{\Gamma_b^2-1}\\
\frac{\rho_b}{\rho_a}&=&\frac{16\Gamma_b^2-10}{6}\left[1+\sqrt{1-{36}~
(16\Gamma^2-10)^{-2}}~\right]\\
M&=&\pm~\frac{n_a\Phi_a(3\rho_b/\rho_a+1)}{4\Gamma_b\sqrt{3\rho_b/\rho_a}}\\
\eta&\equiv&\frac{n_b}{n_a}=\frac{4\Gamma_b}{3(\rho_b/\rho_a)^{-1}+1}\\
\frac{\ep_b}{\ep_a}&=&\frac{3+\rho_b/\rho_a}{4\Gamma_b}\\
\label{eq:phibr}
\Phi_b&=&\frac{\ep_a}{\ep_b}\Phi_a=\frac{3+(\rho_b/\rho_a)^{-1}}{4\Gamma_b}
\Phi_a.
\ea
We note from Eqn~({\ref{eq:phibr}}) that $[\Phi w]=0$.

\centerline{\bf{Appendix D:  Nonrelativistic Shock Tube Problem}}
\@addtoreset{equation}{section}
\def\ksection{\arabic{section}}
\def\theequation{D.\arabic{equation}}
\setcounter{equation}{0}

In this section, we sketch the derivation of
the slab shock tube solution
for nonrelativistic fluids.$^{\ref{Sod}}$
We start with a fluid in which the pressure $p$ and specific volume $V\equiv
1/n$ (where $n$ is the number density)
to the left and right of $x_0$ are $p_1,~V_1$ and $p_2,~V_2$
resectively---thus,
$p$ and $V$ are discontinuous across $x=x_0$.
We assume here that our coordinates are oriented so that $p_1<p_2$ and
$V_1>V_2$.
In addition, the fluid is initially at rest: $v_1=v_2=0$.
Because there is no scale to the problem, the solution is a function of $x/t$
only, where $x$ is the Eulerian coordinate.
At time $t$ later, there are 4 regions.  Region 1 and 2 are at rest
with $p=p_1, V=V_1~{\rm and}~v=v_1$, and $p=p_2, V=V_2~{\rm and}~v=v_2$
respectively.
Separating region 1 and 3 is a shock wave, with position $x_d$.
Region 3 contains the fluid behind the shock, and region 3$'$ is wedged between
regions 3 and 2 and contains the rarefaction wave.  The boundary between
region 3 and 3$'$ is called a contact discontinuity and is at position $x_c$.
The velocity and pressure are constant across this boundary:
$p_{3'}=p_3$ and $v_{3'}=v_3$.  The number density, however, is not.
Finally the boundary between region 3$'$ and region 2 is located at $x_a$.
We will calculate $p,~V~{\rm and~} v$ for regions 3 and 3$'$ at time $t$,
assuming
that $\ep=pV/(\gamma-1)$.  We will solve for the following unknowns:
$p_3,~v_3,~n_3,~n_{3'}$ and $p,~n,~{\rm and~}v$ in the rarefaction wave.

Across a shock front, $v_1-v_3 = \sqrt{(p_3-p_1)(V_1-V_3)}$ and
$\epsilon_1 - \epsilon_3 +\frac{1}{2}(V_1-V_3)(p_1+p_3) = 0$.
Using $\ep=pV/(\gamma-1)$, they become
\ba
\label{eq:V3}
V_3 &=& V_1 [(\gamma+1)p_1 +(\gamma-1)p_3]/[(\gamma-1)p_1 +(\gamma+1)p_3]\\
\label{eq:st}
v_3 &=& -(p_3-p_1)\sqrt{2V_1/[(\gamma-1)p_1 +(\gamma+1)p_3]}
\ea
The similarity solution for a rarefaction wave is $^{\ref{LLcfluids}}$
\ba
\label{eq:prar}
p&=&p_2[1-(\gamma-1)|v|/(2c_2)]^{2/(\gamma-1)}\\
\label{eq:Vrar}
V&=&V_2(p_2/p))^{1/\gamma}\\
\label{eq:vrar}
v &=& 2(c-c_2)/(\gamma-1)\\
\label{eq:absvrar}
|v| &=& 2(c_2-x/t)/(\gamma+1),
\ea
where the local speed of sound is given by $c_i = \sqrt{\gamma p_i V_i}$.
We can equate the speed of the fluid $v_3$ to the fluid velocity
in the rarefaction wave between regions 3 and 3$'$.  We find
\be
\label{eq:stst}
v_{3'} = 2c_2[(p_2/p_3)^{(1-\gamma)/(2\gamma)}-1]/(\gamma-1)
\ee
Combining Eqns~({\ref{eq:st}}) and ({\ref{eq:stst}}), we find that
\be
p_3= p_1 +\frac{1}{\gamma-1} \sqrt{\frac{2\gamma p_2 n_1}{n_2}}
\sqrt{(\gamma-1)
p_1 + (\gamma+1) p_3}\left[1-(p_3/p_2)^{(\gamma-1)(2\gamma)}\right].
\ee
This equation can now be solved numerically for $p_3$.  We can then determine
$v_3,~n_3,~{\rm and~}n_{3'}$ using Eqns~({\ref{eq:st}}), ({\ref{eq:V3}})
and ({\ref{eq:vrar}}).  In addition, we find $p,~n~{\rm and~}v$ from
Eqns~({\ref{eq:prar}}), ({\ref{eq:Vrar}}) and ({\ref{eq:vrar}}).
in the rarefaction wave.

The final step to the solution is
determining the location of the boundaries separating these regions.
Since the rarefaction wave is moving at the sound speed in region 2,
$x_a = tc_2$.
Since the velocity at $b$ is $v_b=v_{3'}$ from Eqn~({\ref{eq:stst}}),
we then use Eqn~({\ref{eq:absvrar}}) to find its position:
$x_b = tc_2 [ 1-(\gamma+1)/(\gamma-1) ~[1-(p_2/p_3)^{(1-\gamma)(2\gamma)} ] ]$.
The contact discontinuity moves with the fluid and its position is thus
$x_c = v_3 ~t$.
Finally, the shock position can be determined by transforming to a frame in
which the shock
is constant and using mass conservation. We then transform back
to find the the velocity of the shock to be
$v_s = ({n_3}/({n_3 -n_1})) v_3$ so that $x_d = v_s t$.

\vspace{.5in}
\centerline{\bf REFERENCES}

\vspace{12pt}

\frenchspacing
\begin{enumerate}

\item\label{Guth} A. Guth, {\em Phys. Rev.} {\bf D 23}, 347 (1981);
	A. Guth and E. Weinberg, {\em Nuc, Phys.}, {\bf B 212}, 321, (1983).

\item\label{TEU} E. W. Kolb and M. S. Turner, {\em The Early Universe},
        (Addison-Wesley, Redwood City, Ca., 1990).

\item\label{LASTEIN} D. La and P. J. Steinhardt, {\em Phys. Rev. Lett.}
	{\bf 62}, 376 (1989); {\em Phys. Lett.} {\bf 220B}, 375 (1989).

\item\label{WatWid}S. Hawking, J. Stewart and I. Moss, {\em Phys. Rev. D}
	{\bf 26}, 2681 (1982); R. Watkins and L. Widrow, {\em Nucl. Phys.}
	{\bf B374}, 446 (1992).

\item\label{GoldZag} Attempts have been made in the
infinitely thin-wall limit to obtain solutions for the evolution of very large
bubbles during extended inflation.
(D. Goldwirth and H. Zaglauer, {\em Phys. Rev. Lett.}, {\bf 67}, 3639, (1991).)
They found that under certain conditions, large
bubbles could collapse.  However, the very assumptions
used in the calculation have been found to be in conflict with the
value of Newton's constant today (A. Liddle and D. Wands, {\em Phys. Rev. D}
{\bf 46}, 3655 (1992)).

\item\label{Wein}E. J. Weinberg, {\em Phys. Rev. D} {\bf 40}, 3950 (1989).

\item\label{BLS} D. La, P. J. Steinhardt, and E. Bertschinger,
         {\em Phys. Lett.} {\bf 231B}, 231 (1989)

\item\label{TurWeiWid} M. Turner, E. Weinberg, L. Widrow,
        {\em Phys. Rev. D} {\bf 46}, 2384 (1992).

\item\label{LiddWan}A. Liddle and D. Wands, {\em Mon. Not. Roy. Aston. Soc.},
	{\bf 253}, 637, (1991).
\item\label{TolBon} R. Tolman, {\em Proc. Natl. Acad. Sci.}, {\bf 20}, 169,
	(1934);
	H. Bondi, {\em Mon. Not. R. Astron. Soc.}, {\bf 107}, 410, (1947).

\item\label{ThinShell} K. Maeda, {\em Gener. Relat and Gravit}, {\bf 18}, 931,
(1986);
\item\label{LLfluids} L. Landau and E. Lifshitz, {\em The Classical Theory
	of Fields}, (Permagon Press, $4^{th}$ edition, 1975).
\item\label{MisSharp}C. Misner and D. Sharp, {\em Phys. Rev }, {\bf 136}, B571,
(1964);
	W. Hernandez and C. Misner, {\em Astro. J.} {\bf 143}, 452, (1966);
	M. May and R. White {\em Phys. Rev}, {\bf 141}, 1232, (1966).
\item\label{MayWhite} M. May and R. White, {\em Meth. Computat. Phys},
	{\bf 73}, 219 (1967).
\item\label{PirBlud} P. Schinder, S. Bludman and T. Piran, {\em Phys. Rev. D},
	{\bf 37}, 2722 (1988).
\item\label{MortRich} R. Richmyer and K. Morton, {\em Difference Methods for
	Initial-Value Problems}, Interscience Publishers, New York(1967).
\item\label{Temp} The temperature can be defined locally since each
	spacetime point is surrounded by a small region which is in
	approximate thermal equilibrium and for which the Bolzmann
	distribution is approximately independent of
	direction.$^{{\ref{SWein}}}$
\item\label{SWein} S. Weinberg, {\em Gravitation and Cosmology}, John
	Wiley \& Sons, New York, (1972).
\item\label{Schutz} B. Schutz, {\em A first course in General Relativity},
	(Cambridge University Press, Cambridge, London, 1985)
\item\label{LLcfluids} L. Landau and E. Lifshitz, {\em Fluid Mechanics},
	Pergamon Press, London, (1959).
\item\label{CourFrei} R. Courant and K. Friedrichs, {\em Supersonic Flow and
	Shock Waves}, Springer-Verlag, New York, (1985).
\item\label{VonNRich} J. VonNeumann and R. Richmyer, {\em Jour. of Applied
Physics},
	{\bf 21}, 232, (1950).
\item\label{BerHobSm} D. Bernstein, D. Hobill and L. Smarr, {\em Dynamical
	Spacetimes and Numerical Relativity ; proceedings}, ed. by J. Centrella,
	Cambridge, New York,  (1986).

\item\label{Comp}This is strictly only true for
	compensated voids, but is also approximately true for uncompensated
	voids for some amount of time.
\item\label{local} Note that in order for the fluid to begin and remain
	in local thermal equilibrium, we must require the interaction
	length of the mediating particle to be
	smaller than the grid spacing at any time $t$.
	If $\lambda$ is the proper interaction length, then
	$\lambda<\Lambda_i^j\Delta r=\Delta R_i^j/\Gamma_i^j$.
	If a gauge boson is the mediating particle (e.g. a photon), then
	$\lambda\simeq (\alpha^2T)^{-1}\simeq g_*^{1/4}/(\alpha^2\rho^{1/4})$,
	where $\sqrt{4\pi\alpha}$ is the gauge coupling,
	$\rho=\pi^2 g_*/30$ and $g_*$ is the number of effectively massless
	degrees of freedom.$^{\ref{TEU}}$

\item\label{Stoneetal} J. Stone et. al., {\em Astrophys. J.},
	{\bf 388}, 415 (1992).

\item\label{Newm} R P A C Newman, {\em Class. Quantum Grav.}, {\bf 3}, 527,
(1986)

\item\label{Bert} E. Bertschinger, {\em Ap.J. Suppl.}, {\bf 58}, 1, (1985)
\item\label{Sod} G. Sod, {\em J. Comput. Physics}, {\bf 27}, 1 (1978).

\item\label{Sedov} Sedov, {\em Similarity and Dimensional Methods in
	Mechanics}, Academic Press, New York, 1959.
\item\label{WallHeat} K. Winkler and M. Norman, {\em Astrophysical
	Radiation Hydrodynamics}, ed. by K. Winkler and M. Norman,
	D. Reidel Publishing Co., Boston, 1986.  (pg. 101).
\item\label{La} D. La, {\em Physics Letters}, {\bf B 265}, 232, (1991).

\item\label{Vasha2} S. Vadas, {\em Numerical Evolution of General Relativistic
Voids II}, work in preparation.
\item\label{Vasha} S. Vadas, {\em Evolution of SuperHorizon-Sized Voids from
	Extended Inflation}, work in preparation.
\item\label{Synge} J. Synge, {\em Relativity: The General Theory},
	North-Holland Publ., Amsterdam (1960).

\end{enumerate}
\nonfrenchspacing
\centerline{\bf FIGURE CAPTIONS}

{\bf Fig.~1:} The evolution of a compensated, pressureless void with radius
$\RW=333c\Hm$ and for velocity $U(\ti,R)=\sqrt{2G_NM/R}$.  The energy density
as a function of $R R(\ti,R_B)/R(t,R_B)\simeq R(\ti/t)^{2/3}$ is shown at the
initial time $\ti=1$ and at times $t=10$, $100$ and $300$.

\vspace{10pt}

{\bf Fig.~2:}  Same as for Figure 1 but with velocity $U(\ti,R)=c^{-1}R
\sqrt{8\pi G_N\rho/3}$.
A dense, thin, outward-moving shell is formed.
Shell-crossing occurs at $t=1.048$.

\vspace{10pt}

{\bf Fig.~3a:} The evolution of a pressureless,
uncompensated void with radius
$\RW=.0067c\Hm$ and velocity $U(\ti,R)=c^{-1}R\sqrt{8\pi G_N\rho/3}$.
Artificial
viscosity is used to prevent shell-crossing.
The energy density
as a function of $R R(\ti,R_B)/R(t,R_B)\simeq R(\ti/t)^{2/3}$ is shown at the
initial time $\ti=1$ and at times $t=50$, $100$, $400$ and $1000$.

\vspace{10pt}

{\bf Fig.~3b:} Same as Figure 3a but for a compensated void.

\vspace{10pt}

{\bf Fig.~3c:} The radius of the shell
versus time for the results pictured in Figures 3a and 3b.  The predicted
self-similar solutions are overlaid as a dotted and dashed line for an
uncompensated ($R_{\rm shell}\propto t^{8/9}$) and compensated ($R_{\rm shell}
\propto t^{4/5}$) void respectively.

\vspace{10pt}

{\bf Fig.~4:} The relative errror in the energy density versus $R R(\ti,R_B)/
R(t,R_B)\simeq R\sqrt{\ti/t}$
for an initially homogeneous and isotropic FRW universe with $p=\rho/3$
at time $\Delta t/\ti=.1$.  The results are
for $\f=.01$, $.005$, and $.0025$ shown with solid, dotted and dashed lines,
respectively.

\vspace{10pt}

{\bf Fig.~5:} The nonrelativistic shock tube for $\RW=20$,
$\rw=.01$ and $G_N=0$.  The dashed lines represent the exact slab similarity
solution.

\vspace{10pt}

{\bf Fig.~6a:} The nonrelativistic evolution of an uncompensated void
with $G_N=0$.
The dashed line is the initial time $\ti=1$, and the triangles with connecting
lines is at time $t=2.5$ {\it after} the inbound shock has bounced and
the fluid has begun to settle down.

\vspace{10pt}

{\bf Fig.~6b:} Same as Figure 6a but for a compensated void.

\vspace{10pt}

{\bf Fig.~7a:} Collapse of a special relativistic, uncompensated void for
$\RW=1$, $\rw=.02$, $T/\mu\gg 1$, $G_N=0$,
$U(\ti,R)=0$ and $\alpha=10^{-4}$.  Shown is $M$, $\Gamma$, $4\pi\rho$
and $\Phi$ versus $R$ for the initial time $\ti=1$ and for times
$t=1.04$ and $1.065$ (before the shock reaches the origin).

\vspace{10pt}

{\bf Fig.~7b:} We plot $4\pi p$ versus $R$ for special relativistic voids
with initial varying wall thicknesses at the initial time (unlabeled)
and at later times $t=1.04$ and $1.07$, $1.08$ or $1.065$.
The parameters are the
same as in Figure 7a but with $\rw=.02$, $.04$, $.06$ and
$.1$.

\vspace{10pt}

{\bf Fig.~7c:}  Collapse, thermalization and homogenization of a special
relativistic void.  The parameters are the same as in Figure 7a.
Shown is the initial time $\ti$ and times
$t=1.04$ (void collapsing) and $t=3.0$ (after collapse).

\vspace{10pt}

{\bf Fig.~8:} Collapse of a general relativistic, uncompensated void for
$\RW=1$, $\rw=.02$, $T/\mu\gg 1$,
$U(\ti,R)=\sqrt{2G_N M/R}$ and $\alpha=10^{-4}$ at the initial time $\ti=1$
and at times $t=1.04$ and $1.065$.

\vspace{10pt}

{\bf Figs.~9a,b,c:}  Collapse of general relativistic, uncompensated voids for
$\RW/$ $\Hm=25$, $\rw=1$, $T/\mu\gg 1$,
$U(\ti,R)=\sqrt{2G_N M/R}$.  Figures 9a, 9b and 9c show the pressure
versus $R R(\ti,R_B)/R(t,R_B)\simeq R\sqrt{\ti/t}$ for
$\alpha=10^{-4}$,
$\alpha=10^{-6}$ and $\alpha=10^{-10}$, respectively.

\vspace{10pt}

{\bf Figs.~10a,b:} Collapse of general relativistic,
uncompensated and compensated void in 10a and b, respectively.
$\RW/\Hm=250$, $\rw=10$, $T/\mu\gg 1$,
$U(\ti,R)=\sqrt{2G_N M/R}$ and $\alpha=10^{-10}$ for both.
Shown is the pressure versus $R R(\ti,R_B)/R(t,R_B)\simeq R\sqrt{\ti/t}$.

\vspace{10pt}

{\bf Fig.~11:} Collapse of a general relativistic, compensated void for
$\RW/\Hm=250$, $\rw=10.$, $T/\mu\gg 1$,
$U(\ti,R)=c^{-1}R\sqrt{8\pi G_N\rho/3}$ and $\alpha=10^{-10}$.
Shown is the pressure versus $R R(\ti,R_B)/R(t,R_B)\simeq R\sqrt{\ti/t}$.

\vspace{10pt}

{\bf Fig.~12:} $1^{\rm st}$-crossing times versus $R\sqrt{\ti/t}$
for general relativistic voids with
initial radius $\RW=c10^{23}\Hm$ and outside temperature $T_{\rm out}=T_{\rm
RH}$,
where $T_{\rm RH}$ is the reheat temperature.
The dash-dot lines are for $\Phi_{\rm in}(\ti)=T_{\rm out}(\ti)/T_{\rm
in}(\ti)$
$=5\times 10^{11}$, $10^{15}$ and $10^{20}$ and
the temperature at which each void potentially thermalizes is
$T=500$, $10^6$ and $10^{11}$GeV respectively.  We also plot (long dashes)
the evolution
of a conformally stretched perturbation with the same initial size, $r_{\rm
CM}$,
and the Hubble radius (short dashes) outside the void, $r_{\rm HOR}$.
If $T_{\rm RH}=10^{14}$GeV, then
recombination occurs at $t/\ti=10^{45}$ (or at temperature
$T_{\rm rec}\simeq 3\times 10^{-9}$GeV), where
$r_{\rm CM}$ and $r_{\rm HOR}$ intersect.

\end{document}